\documentclass[pra,twocolumn,showpacs,superscriptaddress]{revtex4}%
\usepackage{amssymb}
\usepackage{amsmath}
\usepackage{graphicx}
\usepackage{dcolumn}
\usepackage{bm}
\usepackage{amsfonts}%
\providecommand{\U}[1]{\protect\rule{.1in}{.1in}}

\begin{document}
\preprint{$\omega_{k^{\prime}}\omega_{k}$APS/123-QED}
\title{Nonlinear coupling of photons via a collective mode of transparent superconductor}
\author{V.\ M.\ Akulin}
\affiliation{Laboratoire Aim\'{e} Cotton, CNRS (UPR 3321), B\^{a}timent 505, 91405 Orsay
Cedex, France. }
\affiliation{Institute for Information Transmission Problems of the Russian Academy of
Science, Bolshoy Karetny per. 19, Moscow, 127994, Russia.}
\affiliation{Laboratoire J.-V. Poncelet CNRS (UMI 2615) Bolshoi Vlassievsky per. 11,
Moscow, 119002 Russia.}

\begin{abstract}
At the first glance, the expression "transparent superconductor" may seem an
oxymoron. Still, the first principle calculations\cite{Nakanishi} and
experiments\cite{Exp} show that the materials that behave as superconductors
at low frequencies and do not absorb in the optical domain may exist. Virtual
excitation of the collective electronic modes of such superconductors in a
magnetic field appears as an efficient way to realize the nonlinear
interaction of light at the level of two single photons. The essence of the
effect is in the fact that the pondermotor energy is proportional to the ratio
of the charge squared to the mass of the "collective particle" interacting
with radiation, $e^{2}/m$, and therefore, for a "particle" representing a
collective motion of many electrons, it scales linearly-, and its second-order
correction quadratically with the number of the electrons involved. This
general situation is analyzed in detail in the framework of a simple model of
a fiber tube waveguide equipped with a clean superconductor layer. It turns
out that for realistic parameters, at the $%
\mu
$-scale of the tube diameter and the $cm$-scale of the fiber length, such a
\ system is capable of performing the logic gate operation on the polarization
variables of a pair of optical photons.

\end{abstract}

\maketitle

Interaction of photons mediated by atomic or condensed-matter electrons gets
stronger when the latter are in a collective or a cooperative\cite{Dicke}
quantum state.\ Excitations of the Cooper pair condensate in superconductors,
whose interaction with photons is well-described since long
ago\cite{Abrikosov},\cite{Kulik},\cite{Klein}, is one of the examples of such
collective states. It will be shown here, that these states in "transparent
superconductors" can mediate a rather strong coupling of a pair of single
photons. It might happen that the optically transparent and superconducting
substance required for the purpose does not yet exist, but can be predicted by
the first principle calculations, like it has been done\cite{Nakanishi} for
one of the candidates --$p$-doped $CuAlO_{2}$. Searching for such materials is
worth to be done in view of the importance of the visible or near infrared
light manipulation at the level of single photons for Quantum Informatics. The
present paper starts with consideration of interaction of photons with
transparent materials followed by calculation of the nonlinear susceptibility
of a "clean" superconductor and detailed for a specific setting of photon
propagation in a tube waveguide enveloping a thin superconducting layer.

The non-relativistic Pauli equation for an electron in an external quantized
electromagnetic field suggests the interaction term in the form$-\frac{e}%
{mc}\widehat{\overrightarrow{p}}\widehat{\overrightarrow{A}}+\frac
{e^{2}\widehat{\overrightarrow{A}}^{2}}{2mc^{2}}$, where the vector potential
operator%
\begin{equation}
\widehat{\overrightarrow{A}}(\overrightarrow{r})=\sum_{k}\sqrt{\frac{c\pi\hbar
v}{\omega_{k}}}\left(  \widehat{a}_{k}\overrightarrow{u}_{k}\left(
\overrightarrow{r}\right)  +\widehat{a}_{k}^{+}\overrightarrow{u}_{k}^{\ast
}\left(  \overrightarrow{r}\right)  \right)  \label{EQ2}%
\end{equation}
is given in terms of photon frequency $\omega_{k}$, the group velocity
$v$\cite{UFN}, and the creation $\widehat{a}_{k}^{+}$ and annihilation
$\widehat{a}_{k}$ operators of the photons with the mode functions
$\overrightarrow{u}_{k}\left(  \overrightarrow{r}\right)  $ normalized by the
volume integral $\int\overrightarrow{u}_{k}\overrightarrow{u}_{k}^{\ast}dV=1$.
The scalar product is implicit.

From the viewpoint of the relativistic Dirac equation, the pondermotor term
$e^{2}\widehat{\overrightarrow{A}}^{2}/2mc^{2}$ containing the square of the
electron-photon interaction divided by the energy of an electron-positron pair
at rest can be interpreted as a second order relativistic perturbation,
usually small, unless the electromagnetic field is really strong, as for the
case of multiphoton laser ionization of atoms\cite{MichaFedorov}. However, for
transparent materials, where no resonant levels are available for optical
transitions, the main term $-\frac{e}{mc}\widehat{\overrightarrow{p}}%
\widehat{\overrightarrow{A}}$ also gives just a second order contribution,
which is yet smaller than $e^{2}\widehat{\overrightarrow{A}}^{2}/2mc^{2}$ by a
factor $\sim$ $v_{F}/c$ - the ratio of the Fermi and the light velocities. See
\ref{Appendix A} Appendix for the details of the estimate.

For a multi-electron system with the electron density $n_{e}$, the interaction
can be written as $n_{e}e^{2}\overrightarrow{A}^{2}/2mc^{2}\equiv\left(
\omega_{p}/\omega_{k}\right)  ^{2}\overrightarrow{E}^{2}/8\pi$, where
$\omega_{p}=\sqrt{4\pi n_{e}e^{2}/m}$ is the plasma frequency. Even for not
absorbing media, the electromagnetic field at frequencies $\omega_{k}%
<\omega_{p}$ can penetrate at most at the length $\lambda_{p}\sim2\pi
\omega_{k}/c\sqrt{\left(  \omega_{p}/\omega_{k}\right)  ^{2}-1}$, which
implies that at least one of the spacial dimensions of the superconductor
should be less than $\lambda_{p}$ for the transparence required. For a pair of
photons at close frequencies $\omega_{k}$ and $\omega_{k^{\prime}}$ far
detuned from two-photon resonances, in the pondermotor interaction term
\begin{equation}
\widehat{n}_{e}\frac{e^{2}\pi\hbar v}{2mc}\sum_{k;k^{\prime}}\frac{\widehat
{a}_{k}\widehat{a}_{k^{\prime}}^{+}\overrightarrow{u}_{k^{\prime}}^{\ast
}\left(  \overrightarrow{r}\right)  \overrightarrow{u}_{k}\left(
\overrightarrow{r}\right)  +h.c.}{\sqrt{\omega_{k^{\prime}}\omega_{k}}},
\label{EQ4}%
\end{equation}
one can retain only the terms oscillating at the photon frequency difference,
which can be tuned close to the resonance with collective modes of the
superconductor. Here $\widehat{n}_{e}=\widehat{\psi}^{\dag}(\overrightarrow
{r})\widehat{\psi}(\overrightarrow{r})$ is the electron density operator given
in terms of the anticommuting electron creation $\widehat{\psi}^{\dag
}(\overrightarrow{r})$ and annihilation $\widehat{\psi}(\overrightarrow{r})$
field operators, and $h.c.$ denotes Hermite conjugate.

Consider now such a system for the case of a superconductor at zero
temperature in a static sub-critical magnetic field given by the vector
potential $\overrightarrow{A}_{st}$. Each of the photons is in a superposition
of longitudinal modes at close frequencies $\omega_{k}\cong\omega$ and
$\omega_{k^{\prime}}\cong\omega+\delta\omega$, respectively, such that
$\hbar\delta\omega$ is less than the gap parameter $\Delta(\overrightarrow
{r})$. The atomic units $m=1$, $\hbar=1$, $e=1$ are employed hereafter for
shortness. In the framework of the model with a local coupling $-g$, the
corresponding Hamiltonian reads
\begin{align}
\widehat{H}  &  =\int dV\left[  \frac{1}{2}\widehat{\psi}_{s}^{\dag
}(\overrightarrow{r})\left(  \widehat{\overrightarrow{p}}-\widehat
{\overrightarrow{A}}_{st}/c\right)  ^{2}\widehat{\psi}_{s}(\overrightarrow
{r})\right. \label{EQ20}\\
&  -\frac{1}{2}\widehat{\Delta}(\overrightarrow{r})\widehat{\psi}_{s}^{\dag
}(\overrightarrow{r})\widehat{\psi}_{-s}^{\dag}(\overrightarrow{r})-\frac
{1}{2}\widehat{\Delta}^{\dag}(\overrightarrow{r})\widehat{\psi}_{s}%
(\overrightarrow{r})\widehat{\psi}_{-s}(\overrightarrow{r})\nonumber\\
&  +\frac{1}{2g}\widehat{\Delta}^{\dag}(\overrightarrow{r})\widehat{\Delta
}(\overrightarrow{r})+\frac{\pi v}{2c}\widehat{\psi}_{s}^{\dag}%
(\overrightarrow{r})\widehat{\psi}_{s}(\overrightarrow{r})\times
\sum_{k;k^{\prime}}\nonumber\\
&  \left.  \frac{\widehat{a}_{k}\widehat{a}_{k^{\prime}}^{+}\overrightarrow
{u}_{k^{\prime}}^{\ast}\left(  \overrightarrow{r}\right)  \overrightarrow
{u}_{k}\left(  \overrightarrow{r}\right)  +h.c.}{\sqrt{\omega_{k^{\prime}%
}\omega_{k}}}\right]  +\sum_{k}\omega_{k}\left(  \widehat{a}_{k}^{+}%
\widehat{a}_{k}+\frac{1}{2}\right) \nonumber
\end{align}
where summation over the spin subscripts $s=\pm$, which denotes $\pm1/2$, is
implicit. Magnetic interaction with spins is ignored.

In the case where photons are out of resonance with the superconductor
excitations, the interaction among them can be considered as the second order
perturbation, such that the photon part of the Hamiltonian adopts the form%
\begin{equation}
\widehat{H}=\sum_{k}\omega_{k}\left(  \widehat{a}_{k}^{+}\widehat{a}_{k}%
+\frac{1}{2}\right)  +\sum_{kk^{\prime}}\chi_{k,k^{\prime},k^{\prime}%
,k}\left(  \delta\omega\right)  \widehat{a}_{k}\widehat{a}_{k^{\prime}}%
^{+}\widehat{a}_{k^{\prime}}\widehat{a}_{k}^{+}, \label{EQ21}%
\end{equation}
which implies that the nonlinear coupling of photons occur via the linear
susceptibility of the multielectronic system to the pondermotor perturbation
$\widehat{\overrightarrow{A}}^{2}/2c^{2}$. This nonlinear susceptibility
$\chi_{k,k^{\prime},k^{\prime},k}$ of the transparent superconductor is the
very quantity to be calculated.

It is convenient to invoke a standard technique -- the Feynmann integration
over the anticommuting electron fields and classical fields for the order
parameter\cite{Kleinert}.\ In the framework of this approach,%
\begin{equation}
\chi_{k,k^{\prime},k^{\prime},k}=\frac{\partial^{2}}{\partial\alpha
\partial\alpha^{\ast}}\ln Z\left(  \alpha^{\ast},\alpha\right)  , \label{EQ26}%
\end{equation}
where the "partition function" $Z$ is given by the functional integral in the
momentum representation%
\begin{equation}
Z=\int e^{iS}D\psi_{+}^{\ast}D\psi_{+}D\psi_{-}^{\ast}D\psi_{-}D\Delta
_{1}^{\ast}D\Delta_{1}D\Delta_{2}^{\ast}D\Delta_{2} \label{EQ27}%
\end{equation}
with the action%
\begin{align}
S  &  =\int\left(
\begin{array}
[c]{cccc}%
\psi_{+}^{\ast} & \psi_{+}^{\ast} & \psi_{-} & \psi_{-}%
\end{array}
\right)  \widehat{M}\left(
\begin{array}
[c]{c}%
\psi_{+}\\
\psi_{+}\\
\psi_{-}^{\ast}\\
\psi_{-}^{\ast}%
\end{array}
\right)  \frac{d\widetilde{\omega}}{2}d^{3}\widetilde{k}\nonumber\\
&  +\int\frac{1}{2g}\delta\Delta^{\ast}(\omega,\overrightarrow{k})\delta
\Delta(\omega,\overrightarrow{\widetilde{k}})d\widetilde{\omega}%
d^{3}\widetilde{k}. \label{EQ28}%
\end{align}
The Hamiltonian Eq.(\ref{EQ20}) corresponds to the matrix $\widehat{M}$ of the
form\cite{Expr}%
\begin{equation}
\left(
\begin{array}
[c]{cccc}%
\delta\omega+\widetilde{\omega}-\text{$\epsilon_{1}$} & \alpha^{\ast
}\overrightarrow{u}_{k}^{\ast}\overrightarrow{u}_{k^{\prime}} & -\Delta
\text{$_{1}$} & -\Delta\\
\alpha\overrightarrow{u}_{k^{\prime}}^{\ast}\overrightarrow{u}_{k} &
\widetilde{\omega}-\text{$\epsilon_{2}$} & -\Delta & -\Delta_{2}\\
-\Delta^{\ast}\text{$_{1}$} & -\text{$\Delta^{\ast}$} & \widetilde{\omega
}+\text{$\epsilon_{3}$} & -\alpha\overrightarrow{u}_{k^{\prime}}^{\ast
}\overrightarrow{u}_{k}\\
-\text{$\Delta^{\ast}$} & -\Delta_{2}^{\ast} & -\alpha^{\ast}\overrightarrow
{u}_{k}^{\ast}\overrightarrow{u}_{k^{\prime}} & \delta\omega+\widetilde
{\omega}+\text{$\epsilon_{4}$}%
\end{array}
\right)  . \label{EQ29}%
\end{equation}
Here $\alpha$ and $\alpha^{\ast}$ replace the operators $\widehat{a}%
_{k}\widehat{a}_{k^{\prime}}^{+}$ and $\widehat{a}_{k^{\prime}}\widehat{a}%
_{k}^{+}$, respectively, $\epsilon_{2},\epsilon_{3}$ and $\epsilon
_{1},\epsilon_{4}$ are energies of electrons comprising Cooper pairs before
and after the photon-induced virtual transition, respectively. The
factors$\ \sqrt{\frac{\pi v}{2c\omega_{k}}}$ are included to the vectors
$\overrightarrow{u}_{k}$ for shortness. The order parameter amplitudes
$\Delta_{1}$, $\Delta_{2}$, and $\Delta$ are specified below. See
\ref{Appendix B} Appendix for the details of the transformations performed.

It is expedient to discuss the gap parameters of Eq.(\ref{EQ29}) in some more
detail. For superconducting systems at zero temperature not interacting with
radiation, the minimum energy attains at a stationary non-zero value of the
gap $\Delta(\overrightarrow{r})$, with a phase dependent on coordinates in the
presence of a non-zero magnetic field switched on before cooling the
conductor. For a small perturbation by virtual absorption of circularly
polarized photons, this variable also may experience a variation $\delta
\Delta=\Delta_{1}(\overrightarrow{r})+\Delta_{2}(\overrightarrow{r})$, which,
in a sense, resembles that of the two-band Leggett model\cite{Leggett},
although it couples the pair's electrons with distinct angular momenta but not
in different conduction bands. This corresponds to a virtual excitation of
non-dissipative collective motion of Cooper pairs in magnetic field at a
frequency below $2\left\vert \Delta\right\vert $, which has resonant structure
and chirality due to the absorbed angular momentum, and thereby drastically
affect tensor $\chi_{k,k^{\prime},k^{\prime},k}$ close to the resonance. For
the virtual transition with no change of the angular momentum, the
perturbation $\Delta_{2}$ turns to coincide (up to a phase of $\Delta$) with
$\Delta_{1}$$^{\ast}$ and cancels the contribution of the latter, such that no
collective virtual excitation occurs.%

\begin{figure}
[h]
\begin{center}
\includegraphics[
height=2.182in,
width=3.1896in
]%
{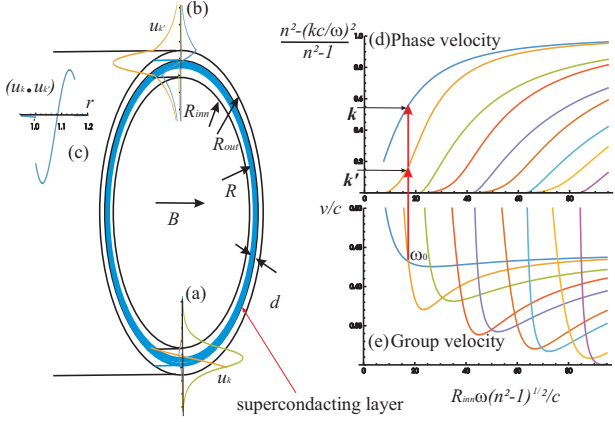}%
\caption{Degenerate tube modes $L=\pm1$ for the refraction index $n=2.26$ and
the ratio of the outer and inner diameters of the tube $R_{out}/R_{inn}%
=1.15$.\ In the inlets (a) The mode function of the first mode. The strong
symmetric component $u_{\theta}$, the asymmetric -$u_{r}$. (b) The second mode
function.\ The strongest symmetric component $u_{r}$, the asymmetric component
$u_{z}$. (c) The scalar product of the mode functions. (d) and (e) The phase
and the group velocities, respectively. On the abscise axis -- scaled
frequency $R_{inn}\frac{\omega}{c}\sqrt{n^{2}-1}$, on the ordinate axis of (d)
-- scaled and shifted phase velocity $\frac{\frac{kc}{\omega}-1}{n-1}$. The
"working point" is near the frequency $\omega_{0}$ where the group velocities
coincide, while the phase velocities and the corresponding wavevectors $k$ and
$k^{\prime}$ are different.}%
\label{FIG1}%
\end{center}
\end{figure}
Consider this situation for a specific setting shown in Fig.\ref{FIG1}.\ The
superconductor is placed as a cylindric layer of radius $R$ and thickness
$d\ll\lambda_{p}$ within a thin wall of a transparent dielectric tube
waveguide with the refraction index $n$. The problem thus has to be formulated
in cylindrical coordinates $r,\theta,z$ and in the momentum representation
$\theta\rightarrow L$, $z\rightarrow k$. The photon mode functions
$\overrightarrow{u}_{k}\left(  \overrightarrow{r}\right)  $ and
$\overrightarrow{u}_{k^{\prime}}\left(  \overrightarrow{r}\right)  $ in the
the axial symmetry setting have the coordinate dependencies
\begin{align}
\sqrt{\frac{\pi v}{2lc\omega_{k}}}\overrightarrow{u}_{k}\left(  r,z,\theta
\right)   &  =\left(
\begin{array}
[c]{c}%
u_{z}\left(  r\right) \\
u_{r}\left(  r\right) \\
u_{\theta}\left(  r\right)
\end{array}
\right)  e^{-i\omega t+ikz+iL\theta}\nonumber\\
& \label{EQ24}\\
\sqrt{\frac{\pi v}{2lc\omega_{k^{\prime}}}}\overrightarrow{u}_{k^{\prime}%
}\left(  r,z,\theta\right)   &  =\left(
\begin{array}
[c]{c}%
u_{z}^{\prime}\left(  r\right) \\
u_{r}^{\prime}\left(  r\right) \\
u_{\theta}^{\prime}\left(  r\right)
\end{array}
\right)  e^{-i\omega^{\prime}t+ik^{\prime}z+iL^{\prime}\theta}\nonumber
\end{align}
that correspond to two modes chosen to have close group velocities $v$ and
$v^{\prime}$. Here, the aforesaid factor $\sqrt{\frac{\pi v}{2c\omega_{k}}}$
also includes the normalization to a waveguide length $l$. The radial mode
functions $u_{z}\left(  r\right)  =-bqZ_{L}\left(  rq\right)  /2k$,
$u_{\theta}\left(  r\right)  =bLZ_{L}\left(  rq\right)  /rq-aZ_{L}^{\prime
}\left(  rq\right)  $, and $u_{r}\left(  r\right)  =aiLZ_{L}\left(  rq\right)
/rq-biZ_{L}^{\prime}\left(  rq\right)  $ are given in terms of the Bessel
functions: $Z_{L}=K_{L}$ outside -, $Z_{L}=I_{L}$ inside - , and $Z_{L}=\gamma
J_{L}+\varkappa Y_{L}$ within the tube wall, and are normalized by the
condition $\int2\pi r\left\vert u\right\vert ^{2}dr=1$. The parameter $q$
amounts to $\sqrt{k^{2}-\omega_{k}^{2}c^{-2}}$ and $\sqrt{n^{2}\omega_{k}%
^{2}c^{-2}-k^{2}}$ outside and inside the wall, respectively. The dispersion
curves $\omega_{k}\left(  k\right)  $ and $\omega_{k^{\prime}}\left(
k^{\prime}\right)  $ correspond to different waveguide modes that are found
numerically from the boundary conditions at the inner and outer radii of the
tube wall. The thin transparent superconducting layer is ignored in the
consideration of the mode fields. \ See \ref{Appendix D} Appendix for the
details of the calculations.

For the electron energies, the cylindrical symmetry implies
\begin{equation}
\text{$\epsilon_{f}$}(\widetilde{L},\widetilde{k})=\frac{p_{r}^{2}}{2}%
+\frac{\widetilde{k}^{2}}{2}+\frac{\left(  \widetilde{L}-\overline{L}\right)
^{2}}{R^{2}}-\mu, \label{EQ25}%
\end{equation}
where $\widetilde{k}$ is the momentum along the axis, $\widetilde{L}$ is the
angular momentum, $p_{r}$ is the radial momentum, $\mu$ is the chemical
potential chosen as the reference energy. The stationary azimuthal magnetic
field potential $\overrightarrow{A}_{st}$ allowing for the magnetic field
parallel to the cylinder axis is parametrized by the number $\overline{L}$ of
the magnetic field quanta traversing the tube cross-section. Note that
$\overline{L}$ accounts here for the magnetic field potential not compensated
by the persistent currents.

If a magnetic field corresponding to $\overline{L}=\Lambda$ was passing
through the waveguide tube cross-section before the layer was cooled down to
the superconducting state, than after the cooling, the magnetic field remains
the same while the stationary order parameter $\Delta$ gets the angular
dependence $e^{i2\Lambda\theta}$ corresponding to zero persistent currents. If
after the cooling, the magnetic field has been further augmented, up to the
value characterized by the parameter $\overline{L}$, the order parameter
angular dependence remains the same, but there appears a persistent current
compensating the augmentation of the magnetic flux through the cross-section,
such that the angular momenta of the electrons now acquire a shift by the
final value of $\overline{L}$. For a thin superconducting layer one can ignore
the radial dependence of $\Delta(\overrightarrow{r})$.

Now one can explicitly find the energies
\begin{align}
\text{$\epsilon_{1}$}  &  =\epsilon_{f}(\widetilde{L}+\delta\text{$L$%
},\widetilde{k}+\delta\text{$k$})\nonumber\\
\text{$\epsilon_{2}$}  &  =\epsilon_{f}(\widetilde{L},\widetilde
{k})\nonumber\\
\text{$\epsilon_{3}$}  &  =\epsilon_{f}(2\Lambda-\widetilde{L},-\widetilde
{k})\nonumber\\
\text{$\epsilon_{4}$}  &  =\epsilon_{f}(2\Lambda-\widetilde{L}-\delta
\text{$L$},-\widetilde{k}-\delta\text{$k$}), \label{EQ30}%
\end{align}
entering Eq.(\ref{EQ29}) and the small perturbation
\begin{equation}
\delta\Delta=\frac{\Delta_{1}e^{iz\delta k-it\delta\omega+i\theta\left(
\delta L+2\Lambda\right)  }+\Delta_{2}e^{-iz\delta k+it\delta\omega
-i\theta\left(  \delta L-2\Lambda\right)  }}{\sqrt{2\pi dRl}}, \label{EQ32}%
\end{equation}
normalized to the layer volume. Here $\delta k=k-k^{\prime}$ and $\delta
L=L-L^{\prime}$ depend on the photon modes and $\Delta_{1,2}$ are the
amplitudes entering Eq.(\ref{EQ29}).

Further a bit cumbersome but completely straightforward calculations can be
sketched as follows. Integration over the anticommuting fields $\psi$ yields%
\begin{equation}
Z=\int e^{\int\frac{d\widetilde{\omega}}{2}\mathrm{Tr}\left[  \log\left(
\det\widehat{M}\right)  \right]  +i\frac{\Delta_{1}^{\ast}\Delta_{1}%
+\Delta_{2}^{\ast}\Delta_{2}}{2g}}d\Delta_{1}\ldots d\Delta_{2}^{\ast},
\label{EQ33}%
\end{equation}
where the first term in the exponent at the right hand side serves as an
action for the variables $\Delta_{1}$ and $\Delta_{2}$ with $\widehat{M}$
given by Eqs.(\ref{EQ29},\ref{EQ30}). After being cast in Taylor series up to
the second order, integrated over the frequency $d\widetilde{\omega}$ and
traced, this term reads%
\begin{equation}
\int d\widetilde{\omega}\mathrm{Tr}\left[  \log\left(  \det\widehat{M}\right)
\right]  \simeq\left(
\begin{array}
[c]{ccc}%
\alpha & \Delta_{1} & \Delta_{2}^{\ast}%
\end{array}
\right)  \widehat{\widetilde{\mathcal{M}}}\left(
\begin{array}
[c]{c}%
\alpha^{\ast}\\
\Delta_{1}^{\ast}\\
\Delta_{2}%
\end{array}
\right)  \label{EQ34}%
\end{equation}
with%
\begin{equation}
\widehat{\widetilde{\mathcal{M}}}=\left(
\begin{array}
[c]{ccc}%
\widetilde{\mathcal{M}}_{\alpha,\alpha} & -\widetilde{\mathcal{M}}%
_{\Delta,\alpha} & \widetilde{\mathcal{M}}_{\Delta,\alpha}\\
-\widetilde{\mathcal{M}}_{\Delta,\alpha} & \widetilde{\mathcal{M}}%
_{\Delta,\Delta} & \widetilde{\mathcal{M}}_{\Delta,\overline{\Delta}}\\
\widetilde{\mathcal{M}}_{\Delta,\alpha} & \widetilde{\mathcal{M}}%
_{\Delta,\overline{\Delta}} & \widetilde{\mathcal{M}}_{\Delta,\Delta}%
\end{array}
\right)  , \label{EQ35}%
\end{equation}
where the matrix elements
\begin{equation}%
\begin{array}
[c]{c}%
\widetilde{\mathcal{M}}_{\alpha,\alpha}=-2i\nu O_{p}\mathcal{I}_{1}\\
\widetilde{\mathcal{M}}_{\Delta,\Delta}=i\nu O_{o}\left(  \mathcal{I}%
_{2}+\mathcal{I}_{4}\right) \\
\widetilde{\mathcal{M}}_{\Delta,\alpha}=-i\nu O_{po}\mathcal{I}_{3}\\
\widetilde{\mathcal{M}}_{\Delta,\overline{\Delta}}=i\nu O_{o}\mathcal{I}_{4}%
\end{array}
\label{EQ41}%
\end{equation}
are given in terms of the integrals%
\begin{equation}%
\begin{array}
[c]{c}%
\mathcal{I}_{1}\left(  \Omega\right)  =\left(  \int\limits_{D\left[  J\right]
}\frac{(\cosh(\xi-\varsigma)+1)d\xi d\varsigma}{\Omega+\cosh\xi+\cosh
\varsigma}\right)  _{+}\\
\mathcal{I}_{2}\left(  \Omega\right)  =\left(  \int\limits_{D\left[  J\right]
}\frac{\left(  -e^{\xi+\varsigma}-1\right)  d\xi d\varsigma}{\Omega+\cosh
\xi+\cosh\varsigma}\right)  _{+}\\
\mathcal{I}_{3}\left(  \Omega\right)  =\left(  \int\limits_{D\left[  J\right]
}\frac{\left(  e^{\xi}+e^{\varsigma}\right)  d\xi d\varsigma}{\Omega+\cosh
\xi+\cosh\varsigma}\right)  _{-}\\
\mathcal{I}_{4}\left(  \Omega\right)  =\left(  \int\limits_{D\left[  J\right]
}\frac{d\xi d\varsigma}{\Omega+\cosh\xi+\cosh\varsigma}\right)  _{+},
\end{array}
\label{EQ45}%
\end{equation}
and where the subscripts $\pm$ denote sum or difference of the integrals in
the parentheses for the positive and the negative scaled frequency $\Omega
=\pm\frac{\delta L(\Lambda-\overline{L})/R^{2}-\Delta\omega}{\left\vert
\Delta\right\vert }$, respectively. The functions $\mathcal{I}_{j}\left(
\Omega\right)  $ diverge logarithmically at the gap edges $\left\vert
\Omega\right\vert =2$ .

The other quantities entering Eq.(\ref{EQ41}) are the mode overlap functions%
\begin{align*}
O_{p}  &  =\frac{\left(  \pi v/c\right)  ^{2}}{2\omega_{k^{\prime}}\omega_{k}%
}\frac{\pi Rd\left(  \overrightarrow{u}_{k}^{\ast}\left(  R\right)
\cdot\overrightarrow{u}_{k^{\prime}}\left(  R\right)  \right)  ^{2}}{l}\\
O_{po}  &  =\frac{\pi v/c}{\sqrt{2\omega_{k}\omega_{k^{\prime}}}}\frac
{\sqrt{\pi Rd}\left(  \overrightarrow{u}_{k}^{\ast}\left(  R\right)
\cdot\overrightarrow{u}_{k^{\prime}}\left(  R\right)  \right)  }{\sqrt{l}}\\
O_{o}  &  =1
\end{align*}
with the restored pre-factors. Since the integrals Eq.(\ref{EQ45}) originate
from the tracing in Eq.(\ref{EQ34}) replaced by the integration over the phase
volume $\mathrm{Tr}\left[  \ldots\right]  \rightarrow\sum_{\widetilde{L}}%
\int\ldots\frac{n_{e}dVdp_{r}d\widetilde{k}}{R\left(  2\pi\right)  ^{3}}$, the
factor
\[
\nu=\frac{\left\vert \text{$\Delta$}\right\vert n_{e}}{8\pi\sqrt
{\frac{\text{$\delta L$}^{2}}{R^{2}}+\text{$\delta k$}^{2}}}%
\]
in Eq.(\ref{EQ41}) is the Jacobian $J$ corresponding to the change of the
phase space integration variables
\[
p_{r},\widetilde{k}\rightarrow\left\{
\begin{array}
[c]{c}%
\xi=\mathrm{arcsinh}\frac{\epsilon_{2}\left(  p_{r},\widetilde{k}\right)
+\epsilon_{3}\left(  p_{r},\widetilde{k}\right)  }{2\left\vert \Delta
\right\vert }\\
\varsigma=\mathrm{arcsinh}\frac{\epsilon_{1\left(  p_{r},\widetilde{k}\right)
}+\epsilon_{4\left(  p_{r},\widetilde{k}\right)  }}{2\left\vert \Delta
\right\vert }%
\end{array}
\right.
\]
summed over the angular momentum and divided by $\cosh\zeta\cosh\xi$ and by
$2\left\vert \Delta\right\vert $ as the result of introducing dimensionless
frequency, while the integration $d\xi d\varsigma$ is restricted to the domain
$D\left[  J\right]  $ where $J$ is real.

The domain $D\left[  J\right]  $ can be explicitly expressed in terms of the
variables $\xi$, $\varsigma$, and two parameters $\widetilde{\mu}=\frac{\mu
}{\left\vert \Delta\right\vert }-\frac{(\overline{L}-\Lambda)^{2}}%
{2R^{2}\left\vert \Delta\right\vert }$ and $\kappa=\frac{\left\vert
\Delta\right\vert }{4\left(  \delta\text{$k^{2}+\delta L^{2}/R^{2}$}\right)
}$. The integration over $\frac{\xi+\varsigma}{2}$ in Eq.(\ref{EQ45}) can be
done analytically yielding a cumbersome but explicit expression dependent on
the other integration variable $\xi-\varsigma$ and these parameters, while the
integration over $\xi-\varsigma$ has to be done numerically. See
\ref{Appendix E} Appendix for the details of the calculations.

After having performed Gaussian integration Eq.(\ref{EQ33}) allowing for
Eq.(\ref{EQ34}), from Eq.(\ref{EQ26}) one obtains the required nonlinear
susceptibility%
\begin{equation}
\chi_{k,k^{\prime},\overline{k}^{\prime},\overline{k}}=\frac{-\left\vert
\text{$\Delta$}\right\vert n_{e}d\left(  \overrightarrow{u}_{k}^{\ast}\left(
R\right)  \cdot\overrightarrow{u}_{k^{\prime}}\left(  R\right)  \right)
^{2}\left(  v/c\right)  ^{2}}{32lR\omega_{k^{\prime}}\omega_{k}\sqrt
{\text{$\delta L$}^{2}/R^{2}+\text{$\delta k$}^{2}}}h\left(  \Omega\right)
\label{EQ43}%
\end{equation}
which couples photons with the wave-vectors $k,k^{\prime},\overline{k}%
^{\prime},\overline{k}$ satisfying the condition $k-k^{\prime}=\overline
{k}-\overline{k}^{\prime}=\delta k$, $L-L^{\prime}=\overline{L}-\overline
{L}^{\prime}=\delta L$. The frequency profile%
\begin{equation}
h\left(  \Omega\right)  =\mathcal{I}_{1}\left(  \Omega\right)  +\frac{\left(
1-\delta_{\delta L}^{0}\right)  \mathcal{I}_{3}^{2}\left(  \Omega\right)
}{\mathcal{I}_{2}\left(  \Omega\right)  +\frac{4\pi\sqrt{\delta\text{$k^{2}%
+\delta L^{2}/R^{2}$}}}{\Delta gn_{e}}} \label{EQ44}%
\end{equation}
is given in terms of the integrals Eq.(\ref{EQ45}) and the Kronekker delta
$\delta_{i}^{j}$, which accounts for the fact that for $\delta L=0$ the
collective amplitudes $\Delta_{1,2}$ coincides (up to a phase of $\Delta$)
with $\Delta_{2,1}^{\ast}$ and give no net contribution.

\begin{figure}
[h]
\begin{center}
\includegraphics[
height=2.0959in,
width=3.2384in
]%
{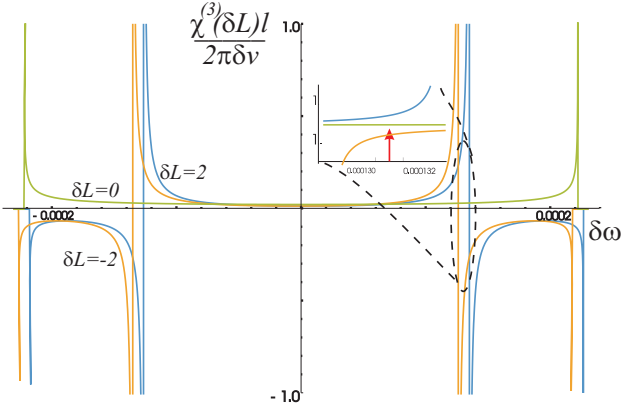}%
\caption{Collective resonance of $\chi_{\delta L}$ in a transparent
superconductor tube of the radius $3$ $\mu$, the critical temperature
$T_{c}=40$ K, the electron density $n_{e}=2.57\times10^{23}$ $cm^{-3}$, the
gap $\Delta=2.2\times10^{-4}$ $\left[  a.u\right]  $, and the thickness $150$
$nm$, which is a typical London length and the radiation penetration depth
$\lambda_{p}$. The chirality of the nonlinear susceptibility is induced by a
constant magnetic field of induction $0.01$ $Ts$ parallel to the tube axis.
The superconductor tube is supported by a tube fiber waveguide with the radii
ratio $R_{out}/R_{inn}=1.15$, and the refraction index $n=2.26$, for the near
infrared light $\lambda=1$ $\mu$. Maximum of the photon vector potentials
scalar product $\left\vert \left(  u_{k}^{\ast}\left(  R\right)  \cdot
u_{k^{\prime}}\left(  R\right)  \right)  \right\vert =0.4$ locates at the
radius corresponding to the superconductor position $R=1.05R_{inn}$. On the
basis axis is the detuning frequency $\delta\omega=\omega-\omega^{\prime}$ in
atomic units, at the ordinate axis the susceptibilities scaled by the factor
$l/2\pi\delta v$ with $\delta v=1.33\times10^{-2}c$ in order to obtain
dimensionless phase shift Eq.(\ref{EQ50}) in fractions of $\pi$ at the
frequency detuning marked by the arrow.}%
\label{FIG2}%
\end{center}
\end{figure}

For a particular case specified in the figure caption, the calculations
Eqs.(\ref{EQ43},\ref{EQ44}) result in the profiles shown in Fig.\ref{FIG2}.%
One sees a strong resonance of the nonlinear susceptibility around the
position of the collective mode, where the denominator in Eq.(\ref{EQ44})
tends to zero. However, such a situation is only possible in a rather narrow
domain of the superconductor parameters where
\begin{equation}
gn_{e}\Delta\lesssim\frac{4\pi\sqrt{\delta\text{$k^{2}+\delta L^{2}/R^{2}$}}%
}{\left\vert \mathcal{I}_{2}\left(  \Omega=0\right)  \right\vert
},\label{EQ45bis}%
\end{equation}
which depends on the superconducting tube radius and the chosen mods.\ Due to
the logarithmic character of the dependence $\mathcal{I}_{2}\left(
\Omega\right)  $ near $\left\vert \Omega\right\vert =2$, the position of the
resonance becomes exponentially close to the band gap edges when the left hand
sides of Eq.(\ref{EQ45bis}) is considerably smaller than the right one. For a
rough estimation with Eq.(\ref{EQ45bis}) one can take $\mathcal{I}_{2}\left(
\Omega=0\right)  \simeq10$. See \ref{Appendix F} Appendix for the details of
the calculations.

For the case of two different degenerate tube modes each of which carries just
a single photon in a linear combination of the left $L=-1$ and the right $L=1$
polarization, the susceptibility is independent of the wave-vectors since
$\delta k=k\left(  \omega\right)  -k^{\prime}\left(  \omega\right)  +O\left(
\frac{\delta\omega}{c}\right)  $ is dominated by the difference of the mode
wave-numbers. For a two-photon state vector%
\begin{equation}
\left\vert \Phi\right\rangle =\sum_{L,L^{\prime}=\pm1}\int dzdz^{\prime}%
\Phi_{L,L^{\prime}}(t,z,z^{\prime})\widehat{a}^{\dag}\left(  z\right)
\widehat{a}^{\prime\dag}\left(  z^{\prime}\right)  \left\vert 0\right\rangle ,
\label{EQ46}%
\end{equation}
given in terms of the operators $\widehat{a}^{\dag}\left(  z\right)  =\sum
_{k}\widehat{a}_{k}^{\dag}e^{-ikz}/\sqrt{l}$ and $\widehat{a}^{\prime\dag
}\left(  z\right)  =\sum_{k}\widehat{a}_{k^{\prime}}^{\dag}e^{ik^{\prime}%
z}/\sqrt{l}$ for the first and the second modes, respectively, the interaction
Hamiltonian of Eq.(\ref{EQ21}) with $\omega_{k}\rightarrow vk$, $\omega
_{k^{\prime}}\rightarrow v^{\prime}k^{\prime}$ results in the Schr\"{o}dinger
equation for the amplitudes $\Phi_{i,j}(t,z,z^{\prime})$%
\begin{align}
0  &  =\left(  i\widehat{\partial}+\chi_{-2}l\delta_{z-z^{\prime}}\right)
\Phi_{-1,-1}(t,z,z^{\prime})\nonumber\\
0  &  =\left(  i\widehat{\partial}+\chi_{2}l\delta_{z-z^{\prime}}\right)
\Phi_{1,1}(t,z,z^{\prime})\nonumber\\
0  &  =\left(  i\widehat{\partial}+\chi_{0}l\delta_{z-z^{\prime}}\right)
\left(  \Phi_{1,-1}(t,z,z^{\prime})+\Phi_{-1,1}(t,z,z^{\prime})\right)
\nonumber\\
0  &  =i\widehat{\partial}\Phi_{1,-1}(t,z,z^{\prime})-i\widehat{\partial}%
\Phi_{-1,1}(t,z,z^{\prime})=0, \label{EQ47}%
\end{align}
where $i\widehat{\partial}=-i\frac{\partial}{\partial t}-iv\frac{\partial
}{\partial z}-iv^{\prime}\frac{\partial}{\partial z^{\prime}}$, $\delta
_{z-z^{\prime}}$ is the Dirac delta function, and the subscript of $\chi$
denotes $\delta L$ of Eq.(\ref{EQ43}).See \ref{Appendix C} Appendix for the
details of the calculations.

The general solution of the equation $\left(  i\widehat{\partial}%
+A\delta_{z-z^{\prime}}\right)  \Phi(t,z,z^{\prime})=0$ reads%
\begin{equation}
\Phi(t,z,z^{\prime})=\Phi(z-vt,z^{\prime}-v^{\prime}t)e^{-i\Theta
_{z-z^{\prime}}\frac{A}{2\delta v}}, \label{EQ48}%
\end{equation}
where $\Theta_{z-z^{\prime}}$ is the Haviside step function and $\delta
v=v-v^{\prime}$. This means that once a photon wave packet of a given circular
polarization in the mode with the higher group velocity overtakes that of the
slower mode, the system acquires the phase shift $\varphi_{\delta L}%
=-\chi_{\delta L}l/2\delta v$ which depends on polarizations of the photons.
For the linear combination of different polarizations in each mode, the
transformation is multiplication $\Phi_{i,j}\left(  t\rightarrow\infty\right)
=U_{i,j}^{k,l}\Phi_{k,l}\left(  t\rightarrow-\infty\right)  $ of the
amplitudes $\left(  \Phi_{-1,-1},\Phi_{1,-1},\Phi_{-1,1},\Phi_{1,1}\right)  $
by the matrix%
\begin{equation}
\widehat{U}=\left(
\begin{array}
[c]{cccc}%
e^{i\varphi_{-2}} & 0 & 0 & 0\\
0 & \frac{e^{i\varphi_{0}}+1}{2} & \frac{e^{i\varphi_{0}}-1}{2} & 0\\
0 & \frac{e^{i\varphi_{0}}-1}{2} & \frac{e^{i\varphi_{0}}+1}{2} & 0\\
0 & 0 & 0 & e^{i\varphi_{2}}%
\end{array}
\right)  . \label{EQ50}%
\end{equation}

Numbers are the most fascinating result of the consideration performed. For
the parameters specified in the caption of Fig.\ref{FIG2}, and the detuning
$\delta\omega=0.033\omega$, one obtains phases in the matrix Eq.(\ref{EQ50}):
$\varphi_{-2}\simeq\pi/2$, $\varphi_{2}\simeq-\pi/2$, $\varphi_{0}<\pi/40$.
This means that $\widehat{U}$ is pretty close to a one of standard quantum
logic gates realized on the photon polarization variables. From the conditions
that the photon wave packets\ of a length $\Delta l\sim1mm$ interact during
the time interval $\Delta l/\delta v$, and that this time interval should be
shorter than time of flight $l/v$, one finds the required tube length
$l\gtrsim45mm$. This quantity can be set to the limit of a few photon pules
length $\Delta l$ by the parameter optimization.

Concluding, one can conjecture that the strong chiral optical nonlinearity is
a common property of the transparent superconductor tubes in magnetic fields
that have the parameters close to the dependence suggested by
Eq.(\ref{EQ45bis}). However, the question of what kind of material can
practically be employed for this purpose is open. The answer implies
exploration of the optical absorption spectra of all known superconducting
substances that, moreover, allow deposition at a supporting transparent
surface as a pure homogeneous layer. It also implies first principle
calculations if the exploration will not yield a suitable result.

I am deeply grateful to Andrey Varlamov for the discussion and his comments.%

\begin{widetext}%

\section{Appendix \label{Appendix A}}

In fact,$\frac{e}{mc}\widehat{\overrightarrow{p}}\widehat{\overrightarrow{A}%
}\frac{1}{\hbar\omega_{k}}$ $\frac{e}{mc}\widehat{\overrightarrow{p}}%
\widehat{\overrightarrow{A}}\sim\frac{\left\langle \widehat{p}\right\rangle
^{2}}{\hbar\omega_{k}}$ $\left(  \frac{e}{mc}\right)  ^{2}\widehat
{\overrightarrow{A}}^{2}$, where $\left\langle \widehat{p}\right\rangle $ is a
typical transition matrix element of the momentum. The matrix element value is
of the order of the Fermi momentum $p_{F}$.\ However, it differs from zero if
the initial and the final states of the electron differ in the momentum by the
momentum of the virtually absorbed photon $k\hbar$. Moreover, the initial
state should belong to the occupied electronic states of the Fermi
distribution, and the final -- to the empty states. The width of the energy
slab which satisfies the latter condition is $\delta E\sim\frac{p_{F}\hbar
k}{m}$, and the relative fraction of the\ slab in the phase space is $\delta
E\frac{d\ln\frac{4}{3}\pi p_{F}^{3}}{d\frac{1}{2m}p_{F}^{2}}$ One therefore
arrives at $\left\langle \widehat{\overrightarrow{p}}\right\rangle ^{2}%
\sim\frac{1}{3}\left\langle \widehat{p}\right\rangle ^{2}\sim$ $\frac{1}%
{3}p_{F}^{2}\times\frac{p_{F}\hbar k}{m}\frac{d\ln\frac{4}{3}\pi p_{F}^{3}%
}{d\frac{1}{2m}p_{F}^{2}}\sim\frac{1}{3}p_{F}^{2}\times2\frac{p_{F}\hbar
k}{2p_{F}}\frac{d\ln\frac{4}{3}\pi p_{F}^{3}}{dp_{F}}$ $\sim p_{F}^{2}%
\times\frac{\hbar k}{p_{F}}\sim p_{F}\hbar k$ and hence$\frac{\left\langle
\widehat{p}\right\rangle ^{2}}{\hbar\omega_{k}}\left(  \frac{e}{mc}\right)
^{2}\widehat{\overrightarrow{A}}^{2}\sim\frac{p_{F}k}{m\omega_{k}}\frac{e^{2}%
}{mc^{2}}\widehat{\overrightarrow{A}}^{2}\sim\frac{v_{F}}{c}\frac{e^{2}%
}{mc^{2}}\widehat{\overrightarrow{A}}^{2}$.

\section{Appendix \label{Appendix B}}

With the help of the anticommutation relations for the field operators the
Hamiltonian is going to be set to the form consistent with Eq.(\ref{EQ28}). In
the momentum representation for the uniform static vector potential
$\overrightarrow{A}_{st}$%
\begin{align*}
\widehat{H}  &  =\int dk{\Large \{}\omega_{k}\left(  \widehat{a}_{k}%
^{+}\widehat{a}_{k}+\frac{1}{2}\right)  +\frac{1}{g}\widehat{\Delta}^{\dag
}(k)\widehat{\Delta}(k)+\frac{1}{2}\widehat{\psi}_{+}^{\dag}(k)\left(
k-\overrightarrow{A}_{st}/c\right)  ^{2}\widehat{\psi}_{+}(k)+\frac{1}%
{2}\widehat{\psi}_{-}^{\dag}(k)\left(  k-\overrightarrow{A}_{st}/c\right)
^{2}\widehat{\psi}_{-}(k)\\
&  +\int dk^{\prime}{\LARGE [}-\frac{1}{2}\widehat{\Delta}(k-k^{\prime
})\left(  \widehat{\psi}_{+}^{\dag}(k)\widehat{\psi}_{-}^{\dag}(-k^{\prime
})-\widehat{\psi}_{-}^{\dag}(k)\widehat{\psi}_{+}^{\dag}(-k^{\prime})\right)
-\frac{1}{2}\widehat{\Delta}^{\dag}(k-k^{\prime})\left(  \widehat{\psi}%
_{+}(k)\widehat{\psi}_{-}(-k^{\prime})-\widehat{\psi}_{-}(k)\widehat{\psi}%
_{+}(-k^{\prime})\right) \\
&  +\int dk^{\prime\prime}dk^{\prime\prime\prime}\delta\left(  k-k^{\prime
}-k^{\prime\prime}+k^{\prime\prime\prime}\right)  \frac{\widehat{a}%
_{k^{\prime\prime}}\widehat{a}_{k^{\prime\prime\prime}}^{+}\overrightarrow
{u}_{k^{\prime\prime}}^{\ast}\overrightarrow{u}_{k^{\prime\prime\prime}}%
+h.c.}{\sqrt{\omega_{k^{\prime\prime\prime}}\omega_{k^{\prime\prime}}}%
}{\LARGE (}\frac{\pi v}{2c}\widehat{\psi}_{+}^{\dag}(k)\widehat{\psi}%
_{+}(k^{\prime})+\frac{\pi v}{2c}\widehat{\psi}_{-}^{\dag}(k)\widehat{\psi
}_{-}(k^{\prime}){\LARGE )]}{\Large \}.}%
\end{align*}
The anticommutation yields%
\begin{align*}
\widehat{H}  &  =\int dk{\Large \{}\omega_{k}\left(  \widehat{a}_{k}%
^{+}\widehat{a}_{k}+\frac{1}{2}\right)  +\frac{1}{g}\widehat{\Delta}^{\dag
}(k)\widehat{\Delta}(k)+\frac{1}{2}\widehat{\psi}_{+}^{\dag}(k)\left(
k-\overrightarrow{A}_{st}/c\right)  ^{2}\widehat{\psi}_{+}(k)+\frac{1}%
{2}\widehat{\psi}_{-}^{\dag}(k)\left(  k-\overrightarrow{A}_{st}/c\right)
^{2}\widehat{\psi}_{-}(k)\\
&  +\int dk^{\prime}{\LARGE [}-\frac{1}{2}\left(  \widehat{\Delta}%
(k-k^{\prime})\widehat{\psi}_{+}^{\dag}(k)\widehat{\psi}_{-}^{\dag}%
(-k^{\prime})-\widehat{\Delta}(k-k^{\prime})\widehat{\psi}_{+}^{\dag
}(-k^{\prime})\widehat{\psi}_{-}^{\dag}(k)\right) \\
&  -\frac{1}{2}\left(  -\widehat{\Delta}^{\dag}(k-k^{\prime})\widehat{\psi
}_{-}(-k^{\prime})\widehat{\psi}_{+}(k)+\widehat{\Delta}^{\dag}(k-k^{\prime
})\widehat{\psi}_{-}(k)\widehat{\psi}_{+}(-k^{\prime})\right) \\
&  +\int dk^{\prime\prime}dk^{\prime\prime\prime}\delta\left(  k-k^{\prime
}-k^{\prime\prime}+k^{\prime\prime\prime}\right)  \frac{\widehat{a}%
_{k^{\prime\prime}}\widehat{a}_{k^{\prime\prime\prime}}^{+}\overrightarrow
{u}_{k^{\prime\prime}}^{\ast}\overrightarrow{u}_{k^{\prime\prime\prime}}%
+h.c.}{\sqrt{\omega_{k^{\prime\prime\prime}}\omega_{k^{\prime\prime}}}%
}{\LARGE (}\frac{\pi v}{2c}\widehat{\psi}_{+}^{\dag}(k)\widehat{\psi}%
_{+}(k^{\prime})-\frac{\pi v}{2c}\widehat{\psi}_{-}(k^{\prime})\widehat{\psi
}_{-}^{\dag}(k){\LARGE )]}{\Large \}.}%
\end{align*}
Now one changes the integration variables $k\rightarrow\overline{k}$,
$k^{\prime}\rightarrow\overline{k}^{\prime}$ in two last components of the
vector
\begin{align*}
\widehat{H}  &  =\int dk{\Large \{}\omega_{k}\left(  \widehat{a}_{k}%
^{+}\widehat{a}_{k}+\frac{1}{2}\right)  +\frac{1}{g}\widehat{\Delta}^{\dag
}(k)\widehat{\Delta}(k)+\frac{1}{2}\widehat{\psi}_{+}^{\dag}(k)\left(
k-\overrightarrow{A}_{st}/c\right)  ^{2}\widehat{\psi}_{+}(k)+\frac{1}%
{2}\widehat{\psi}_{-}^{\dag}(k)\left(  k-\overrightarrow{A}_{st}/c\right)
^{2}\widehat{\psi}_{-}(k)\\
&  +\int dk^{\prime}{\LARGE [}-\frac{1}{2}\left(  \widehat{\Delta}%
(k-\overline{k}^{\prime})\widehat{\psi}_{+}^{\dag}(k)\widehat{\psi}_{-}^{\dag
}(-\overline{k}^{\prime})+\widehat{\Delta}(k^{\prime}-\overline{k}%
)\widehat{\psi}_{+}^{\dag}(k^{\prime})\widehat{\psi}_{-}^{\dag}(-\overline
{k})\right) \\
&  -\frac{1}{2}\left(  -\widehat{\Delta}^{\dag}(k-\overline{k}^{\prime
})\widehat{\psi}_{-}(-\overline{k}^{\prime})\widehat{\psi}_{+}(k)-\widehat
{\Delta}^{\dag}(k^{\prime}-\overline{k})\widehat{\psi}_{-}(-\overline
{k})\widehat{\psi}_{+}(k^{\prime})\right) \\
&  +\int dk^{\prime\prime}dk^{\prime\prime\prime}\delta\left(  k-k^{\prime
}-k^{\prime\prime}+k^{\prime\prime\prime}\right)  \frac{\widehat{a}%
_{k^{\prime\prime}}\widehat{a}_{k^{\prime\prime\prime}}^{+}\overrightarrow
{u}_{k^{\prime\prime}}^{\ast}\overrightarrow{u}_{k^{\prime\prime\prime}}%
+h.c.}{\sqrt{\omega_{k^{\prime\prime\prime}}\omega_{k^{\prime\prime}}}%
}{\LARGE (}\frac{\pi v}{2c}\widehat{\psi}_{+}^{\dag}(k)\widehat{\psi}%
_{+}(k^{\prime})-\frac{\pi v}{2c}\widehat{\psi}_{-}(k^{\prime})\widehat{\psi
}_{-}^{\dag}(k){\LARGE )]}{\Large \},}%
\end{align*}
and arrives to the matrix form
\begin{align*}
&  \left(
\begin{array}
[c]{cccc}%
\widehat{\psi}_{+}^{\dag}(k^{\prime}) & \widehat{\psi}_{+}^{\dag}(k) &
\widehat{\psi}_{-}(-\overline{k}) & \widehat{\psi}_{-}(-\overline{k}^{\prime})
\end{array}
\right)  \times\\
&  \left(
\begin{array}
[c]{cccc}%
\frac{1}{2}\left(  k^{\prime}-\overrightarrow{A}_{st}/c\right)  ^{2} &
\frac{\pi v}{2c}\frac{\overrightarrow{u}_{k^{\prime\prime}}^{\ast
}\overrightarrow{u}_{k^{\prime\prime\prime}}\widehat{a}_{k^{\prime\prime}%
}\widehat{a}_{k^{\prime\prime\prime}}^{+}}{\sqrt{\omega_{k^{\prime\prime
\prime}}\omega_{k^{\prime\prime}}}} & \frac{1}{2}\widehat{\Delta}(k^{\prime
}-\overline{k}) & \frac{1}{2}\widehat{\Delta}(S)\\
\frac{\pi v}{2c}\frac{\overrightarrow{u}_{k^{\prime\prime}}\overrightarrow
{u}_{k^{\prime\prime\prime}}^{\ast}\widehat{a}_{k^{\prime\prime}}^{+}%
\widehat{a}_{k^{\prime\prime\prime}}}{\sqrt{\omega_{k^{\prime\prime\prime}%
}\omega_{k^{\prime\prime}}}} & \frac{1}{2}\left(  k-\overrightarrow{A}%
_{st}/c\right)  ^{2} & \frac{1}{2}\widehat{\Delta}(S) & \frac{1}{2}%
\widehat{\Delta}(k-\overline{k}^{\prime})\\
\frac{1}{2}\widehat{\Delta}^{\dag}(k^{\prime}-\overline{k}) & \frac{1}%
{2}\widehat{\Delta}^{\dag}(S) & -\frac{1}{2}\left(  -\overline{k}%
-\overrightarrow{A}_{st}/c\right)  ^{2} & -\frac{\pi v}{2c}\frac
{\overrightarrow{u}_{k^{\prime\prime}}\overrightarrow{u}_{k^{\prime
\prime\prime}}^{\ast}\widehat{a}_{k^{\prime\prime}}^{+}\widehat{a}%
_{k^{\prime\prime\prime}}}{\sqrt{\omega_{k^{\prime\prime\prime}}%
\omega_{k^{\prime\prime}}}}\\
\frac{1}{2}\widehat{\Delta}^{\dag}(S) & \frac{1}{2}\widehat{\Delta}^{\dag
}(k-\overline{k}^{\prime}) & -\frac{\pi v}{2c}\frac{\overrightarrow
{u}_{k^{\prime\prime}}^{\ast}\overrightarrow{u}_{k^{\prime\prime\prime}%
}\widehat{a}_{k^{\prime\prime}}\widehat{a}_{k^{\prime\prime\prime}}^{+}}%
{\sqrt{\omega_{k^{\prime\prime\prime}}\omega_{k^{\prime\prime}}}} & -\frac
{1}{2}\left(  -\overline{k}^{\prime}-\overrightarrow{A}_{st}/c\right)  ^{2}%
\end{array}
\right)  \left(
\begin{array}
[c]{c}%
\widehat{\psi}_{+}(k^{\prime})\\
\widehat{\psi}_{+}(k)\\
\widehat{\psi}_{-}^{\dag}(-\overline{k})\\
\widehat{\psi}_{-}^{\dag}(-\overline{k^{\prime}})
\end{array}
\right)
\end{align*}
for the fermionic part. Now the replacement $\overline{k}^{\prime}\rightarrow
k^{\prime}+S$, $\overline{k}\rightarrow k+S$
\begin{align*}
&  \omega_{k}\left(  \widehat{a}_{k}^{+}\widehat{a}_{k}+\frac{1}{2}\right)
-\frac{1}{2g}\widehat{\Delta}^{\dag}(\overrightarrow{k})\widehat{\Delta
}(\overrightarrow{k})+\left(
\begin{array}
[c]{cccc}%
\widehat{\psi}_{+}^{\dag}(k^{\prime}) & \widehat{\psi}_{+}^{\dag}(k) &
\widehat{\psi}_{-}(-k-S) & \widehat{\psi}_{-}(-k^{\prime}-S)
\end{array}
\right)  \times\\
&  \left(
\begin{array}
[c]{cccc}%
\frac{1}{2}\left(  k^{\prime}-A_{st}/c\right)  ^{2} & \frac{\pi v}{2c}%
\frac{\overrightarrow{u}_{k^{\prime\prime}}^{\ast}\overrightarrow
{u}_{k^{\prime\prime\prime}}\widehat{a}_{k^{\prime\prime}}\widehat
{a}_{k^{\prime\prime\prime}}^{+}}{\sqrt{\omega_{k^{\prime\prime\prime}}%
\omega_{k^{\prime\prime}}}} & \frac{1}{2}\widehat{\Delta}(k^{\prime}-k-S) &
\frac{1}{2}\widehat{\Delta}(S)\\
\frac{\pi v}{2c}\frac{\overrightarrow{u}_{k^{\prime\prime}}\overrightarrow
{u}_{k^{\prime\prime\prime}}^{\ast}\widehat{a}_{k^{\prime\prime}}^{+}%
\widehat{a}_{k^{\prime\prime\prime}}}{\sqrt{\omega_{k^{\prime\prime\prime}%
}\omega_{k^{\prime\prime}}}} & \frac{1}{2}\left(  k-A_{st}/c\right)  ^{2} &
\frac{1}{2}\widehat{\Delta}(S) & -\frac{1}{2}\widehat{\Delta}(k-k^{\prime
}-S)\\
-\frac{1}{2}\widehat{\Delta}^{\dag}(k^{\prime}-k-S) & \frac{1}{2}%
\widehat{\Delta}^{\dag}(S) & -\frac{1}{2}\left(  -k-S-\overrightarrow{A}%
_{st}/c\right)  ^{2} & -\frac{\pi v}{2c}\frac{\overrightarrow{u}%
_{k^{\prime\prime}}\overrightarrow{u}_{k^{\prime\prime\prime}}^{\ast}%
\widehat{a}_{k^{\prime\prime}}^{+}\widehat{a}_{k^{\prime\prime\prime}}}%
{\sqrt{\omega_{k^{\prime\prime\prime}}\omega_{k^{\prime\prime}}}}\\
\frac{1}{2}\widehat{\Delta}^{\dag}(S) & \frac{1}{2}\widehat{\Delta}^{\dag
}(k-k^{\prime}-S) & -\frac{\pi v}{2c}\frac{\overrightarrow{u}_{k^{\prime
\prime}}^{\ast}\overrightarrow{u}_{k^{\prime\prime\prime}}\widehat
{a}_{k^{\prime\prime}}\widehat{a}_{k^{\prime\prime\prime}}^{+}}{\sqrt
{\omega_{k^{\prime\prime\prime}}\omega_{k^{\prime\prime}}}} & -\frac{1}%
{2}\left(  -k^{\prime}-S-\overrightarrow{A}_{st}/c\right)  ^{2}%
\end{array}
\right)  \left(
\begin{array}
[c]{c}%
\widehat{\psi}_{+}(k^{\prime})\\
\widehat{\psi}_{+}(k)\\
\widehat{\psi}_{-}^{\dag}(-k-S)\\
\widehat{\psi}_{-}^{\dag}(-k^{\prime}-S)
\end{array}
\right)  ,
\end{align*}
where $S$ denotes the momentum shift in the presence of a magnetic field.

For the photon with the wavenumber difference $k^{\prime\prime\prime
}-k^{\prime\prime}\rightarrow\delta k$, make replacement in the electron
arguments $k\rightarrow\widetilde{k}$, $k^{\prime}\rightarrow\widetilde
{k}+\delta k$ and the photon arguments $k^{\prime\prime\prime}\rightarrow
k^{\prime}\rightarrow k+\delta k$, $k^{\prime\prime}\rightarrow k$, then the
integrand adopts the form
\begin{align*}
&  \frac{1}{2g}\widehat{\Delta}^{\dag}(\delta k-S)\widehat{\Delta}(\delta
k-S)+\frac{1}{2g}\widehat{\Delta}^{\dag}(-\delta k-S)\widehat{\Delta}(-\delta
k-S)+\omega_{k}\left(  \widehat{a}_{k}^{+}\widehat{a}_{k}+\frac{1}{2}\right)
+\omega_{k+\delta k}\left(  \widehat{a}_{k+\delta k}^{+}\widehat{a}_{k+\delta
k}+\frac{1}{2}\right) \\
&  +\frac{1}{2}\left(
\begin{array}
[c]{cccc}%
\widehat{\psi}_{+}^{\dag}(\widetilde{k}+\delta k) & \widehat{\psi}_{+}^{\dag
}(\widetilde{k}) & \widehat{\psi}_{-+}(-\widetilde{k}-S) & \widehat{\psi}%
_{-+}(-\widetilde{k}-\delta k-S)
\end{array}
\right)  \times\\
&  \left(
\begin{array}
[c]{cccc}%
\left(  \widetilde{k}+\delta k-\frac{A_{st}}{c}\right)  ^{2} & \frac{\pi v}%
{c}\frac{\overrightarrow{u}_{k}^{\ast}\overrightarrow{u}_{k+\delta k}%
\widehat{a}_{k}\widehat{a}_{k+\delta k}^{+}}{\sqrt{\omega_{k+\delta k}%
\omega_{k}}} & \widehat{\Delta}(\delta k-S) & \widehat{\Delta}(S)\\
\frac{\pi v}{c}\frac{\overrightarrow{u}_{k}\overrightarrow{u}_{k+\delta
k}^{\ast}\widehat{a}_{k}^{+}\widehat{a}_{k+\delta k}}{\sqrt{\omega_{k+\delta
k}\omega_{k}}} & \left(  \widetilde{k}-\frac{A_{st}}{c}\right)  ^{2} &
\widehat{\Delta}(S) & \widehat{\Delta}(-\delta k-S)\\
\widehat{\Delta}^{\dag}(\delta k-S) & \widehat{\Delta}^{\dag}(S) & -\left(
-\widetilde{k}-S-\frac{A_{st}}{c}\right)  ^{2} & -\frac{\pi v}{c}%
\frac{\overrightarrow{u}_{k}\overrightarrow{u}_{k+\delta k}^{\ast}\widehat
{a}_{k}^{+}\widehat{a}_{k+\delta k}}{\sqrt{\omega_{k+\delta k}\omega_{k}}}\\
\widehat{\Delta}^{\dag}(S) & \widehat{\Delta}^{\dag}(-\delta k-S) & -\frac{\pi
v}{c}\frac{\overrightarrow{u}_{k}^{\ast}\overrightarrow{u}_{k+\delta
k}\widehat{a}_{k}\widehat{a}_{k+\delta k}^{+}}{\sqrt{\omega k+\delta
k\omega_{k}}} & -\left(  -\widetilde{k}-\delta k-S-\frac{A_{st}}{c}\right)
^{2}%
\end{array}
\right)  \left(
\begin{array}
[c]{c}%
\widehat{\psi}_{+}(\widetilde{k}+\delta k)\\
\widehat{\psi}_{+}(\widetilde{k})\\
\widehat{\psi}_{-}^{\dag}(-\widetilde{k}-S)\\
\widehat{\psi}_{-}^{\dag}(-\widetilde{k}-\delta k-S)
\end{array}
\right)  .
\end{align*}

Before the cooling, the magnetic field potential is given as $\frac{A_{st}}%
{c}\rightarrow\Lambda$, and after the cooling the coupling occur among the
electron states with the same energy $\left(  \widetilde{k}-\Lambda\right)
^{2}=\left(  -\widetilde{k}-S-\Lambda\right)  ^{2}$, hence $S=-2\Lambda$. One
therefore has
\begin{align}
&  \frac{1}{2g}\widehat{\Delta}^{\dag}(\delta k-S)\widehat{\Delta}(\delta
k-S)+\frac{1}{2g}\widehat{\Delta}^{\dag}(-\delta k-S)\widehat{\Delta}(-\delta
k-S)+\omega_{k}\left(  \widehat{a}_{k}^{+}\widehat{a}_{k}+\frac{1}{2}\right)
+\omega_{k+\delta k}\left(  \widehat{a}_{k+\delta k}^{+}\widehat{a}_{k+\delta
k}+\frac{1}{2}\right) \label{EQHamBis}\\
&  +\frac{1}{2}\left(
\begin{array}
[c]{cccc}%
\widehat{\psi}_{+}^{\dag}(\widetilde{k}+\delta k) & \widehat{\psi}_{+}^{\dag
}(\widetilde{k}) & \widehat{\psi}_{-}(-\widetilde{k}-S) & \widehat{\psi}%
_{-}(-\widetilde{k}-\delta k-S)
\end{array}
\right)  \times\nonumber\\
&  \left(
\begin{array}
[c]{cccc}%
\left(  \widetilde{k}+\delta k-\frac{A_{st}}{c}\right)  ^{2} & \frac{\pi v}%
{c}\frac{\overrightarrow{u}_{k}^{\ast}\overrightarrow{u}_{k+\delta k}%
\widehat{a}_{k}\widehat{a}_{k+\delta k}^{+}}{\sqrt{\omega_{k+\delta k}%
\omega_{k}}} & \widehat{\Delta}(\delta k+2\Lambda) & \widehat{\Delta
}(-2\Lambda)\\
\frac{\pi v}{c}\frac{\overrightarrow{u}_{k}\overrightarrow{u}_{k+\delta
k}^{\ast}\widehat{a}_{k}^{+}\widehat{a}_{k+\delta k}}{\sqrt{\omega_{k+\delta
k}\omega_{k}}} & \left(  \widetilde{k}-\frac{A_{st}}{c}\right)  ^{2} &
\widehat{\Delta}(-2\Lambda) & \widehat{\Delta}(-\delta k+2\Lambda)\\
\widehat{\Delta}^{\dag}(\delta k+2\Lambda) & \widehat{\Delta}^{\dag}%
(-2\Lambda) & -\left(  -\widetilde{k}+2\Lambda-\frac{A_{st}}{c}\right)  ^{2} &
-\frac{\pi v}{c}\frac{\overrightarrow{u}_{k}\overrightarrow{u}_{k+\delta
k}^{\ast}\widehat{a}_{k}^{+}\widehat{a}_{k+\delta k}}{\sqrt{\omega_{k+\delta
k}\omega_{k}}}\\
\widehat{\Delta}^{\dag}(-2\Lambda) & \widehat{\Delta}^{\dag}(-\delta
k+2\Lambda) & -\frac{\pi v}{c}\frac{\overrightarrow{u}_{k}^{\ast
}\overrightarrow{u}_{k+\delta k}\widehat{a}_{k}\widehat{a}_{k+\delta k}^{+}%
}{\sqrt{\omega k+\delta k\omega_{k}}} & -\left(  -\widetilde{k}-\delta
k+2\Lambda-\frac{A_{st}}{c}\right)  ^{2}%
\end{array}
\right)  \left(
\begin{array}
[c]{c}%
\widehat{\psi}_{+}(\widetilde{k}+\delta k)\\
\widehat{\psi}_{+}(\widetilde{k})\\
\widehat{\psi}_{-}^{\dag}(-\widetilde{k}+2\Lambda)\\
\widehat{\psi}_{-}^{\dag}(-\widetilde{k}-\delta k+2\Lambda)
\end{array}
\right) \nonumber
\end{align}
This expression implies that after the cooling the magnetic field has been
changed and now it is given by the vector potential $\frac{A_{st}}{c}$, which
is different from $\Lambda$. To avoid confusion note, that later on, for the
case of the cylindric setting, the field vector potential $A_{st}$ will be
parametrized by the number of the magnetic field quanta traversing the
cylinder cross-section and will be treated as an angular momentum.

Also note, that the main role of the "frozen" part of the magnetic potential
$\Lambda$ is to avoid interference $\widehat{\Delta}(\delta k+2\Lambda)$ and
$\widehat{\Delta}(-\delta k+2\Lambda)$, since otherwise, $\widehat{\Delta}%
^{+}(\delta k)=\widehat{\Delta}(-\delta k)$, and the cross couplings of the
Cooper pair's electrons before and after the virtual transition may, and do
cancel the momentum sensitive part of the nonlinear coupling.

The matrix Eq.(\ref{EQ29}) comes from the electron part of the action
Eq.(\ref{EQ28}) with $\widehat{M}=i\partial_{t}-\widehat{H}$, where the
Hamiltonian corresponds to the electron part of Eq.(\ref{EQHamBis}). The time
derivative part experience no transformation when the electron field operators
are interchanged, since sign change due to the change of the order of the
fermionic operators is followed by transferring of the time derivative
operator from the left field operator $\psi$ to the right one, and hence in
the Fourier representation
\[
i\partial_{t}\rightarrow\left(
\begin{array}
[c]{cccc}%
\delta\omega+\widetilde{\omega} & 0 & 0 & 0\\
0 & \widetilde{\omega} & 0 & 0\\
0 & 0 & \widetilde{\omega} & 0\\
0 & 0 & 0 & \delta\omega+\widetilde{\omega}%
\end{array}
\right)  ,
\]
which only allows for the energy shift $\delta\omega$ of the Cooper pair after
virtual absorption of the photon: before the absorption -- positions $2 $ and
$3$, and after the absorption -- positions $1$ and $4$. This form is
consistent with Eq.(4.5) of \cite{Kleinert}

\section{Appendix \label{Appendix D}}

The relations between the magnetic field and the vector potential components
in cylindrical coordinates read%
\begin{align*}
B_{r}  &  =\frac{1}{r}\frac{\partial}{\partial\theta}A_{z}-\frac{\partial
}{\partial z}A_{\theta},\\
B_{\theta}  &  =\frac{\partial}{\partial z}A_{r}-\frac{\partial}{\partial
r}A_{z},\\
B_{z}  &  =\frac{1}{r}\frac{\partial}{\partial r}\left(  rA_{\theta}\right)
-\frac{\partial}{r\partial\theta}A_{r}.
\end{align*}
Outside the tube the fields satisfying the wave equation are%
\begin{align*}
A_{r}  &  =\left[  \frac{i}{2}\overline{D}\frac{m}{rq}K_{m}\left(  rq\right)
-\frac{i}{2}DK_{m}^{\prime}\left(  rq\right)  \right]  e^{i\left(  \omega
t-kz-m\theta\right)  }\\
A_{\theta}  &  =\left[  \frac{1}{2}D\frac{m}{rq}K_{m}\left(  rq\right)
-\frac{1}{2}\overline{D}K_{m}^{\prime}\left(  rq\right)  \right]  e^{i\left(
\omega t-kz-m\theta\right)  }\\
A_{z}  &  =\frac{-q}{2k}DK_{m}\left(  rq\right)  e^{i\left(  \omega
t-kz-m\theta\right)  }\\
B_{r}  &  =\left[  \frac{i}{2}D\frac{\omega^{2}}{k}\frac{m}{rq}K_{m}\left(
rq\right)  -\frac{i}{2}\overline{D}kK_{m}^{\prime}\left(  rq\right)  \right]
e^{i\left(  \omega t-kz-m\theta\right)  }\\
B_{\theta}  &  =\left[  \frac{1}{2}\overline{D}\frac{km}{rq}K_{m}\left(
rq\right)  -\frac{1}{2}\frac{\omega^{2}}{k}DK_{m}^{\prime}\left(  rq\right)
\right]  e^{i\left(  \omega t-kz-m\theta\right)  }\\
B_{z}  &  =-\frac{q}{2}\overline{D}K_{m}\left(  rq\right)  e^{i\left(  \omega
t-kz-m\theta\right)  }%
\end{align*}
where $q=\sqrt{k^{2}-\left(  \frac{\omega}{c}\right)  ^{2}}$, $K_{m}\left(
x\right)  $ are the modified Bessel function regular at $x\rightarrow\infty$,
and $D$, $\overline{D}$ are the constants to be determined. Inside the tube,
for the radial parts multiplying the phase factor $e^{i\left(  \omega
t-kz-m\theta\right)  }$ one takes
\begin{align*}
A_{r}  &  =\frac{i}{2}\overline{A}\frac{m}{rq}I_{m}\left(  rq\right)
-\frac{i}{2}AI_{m}^{\prime}\left(  rq\right) \\
A_{\theta}  &  =\frac{1}{2}A\frac{m}{rq}I_{m}\left(  rq\right)  -\frac{1}%
{2}\overline{A}I_{m}^{\prime}\left(  rq\right) \\
A_{z}  &  =\frac{-q}{2k}AI_{m}\left(  rq\right) \\
B_{r}  &  =\frac{i}{2}A\frac{\omega^{2}}{k}\frac{m}{rq}I_{m}\left(  rq\right)
-\frac{i}{2}\overline{A}kI_{m}^{\prime}\left(  rq\right) \\
B_{\theta}  &  =\frac{1}{2}\overline{A}\frac{km}{rq}I_{m}\left(  rq\right)
-\frac{1}{2}A\frac{\omega^{2}}{k}I_{m}^{\prime}\left(  rq\right) \\
B_{z}  &  =-\frac{q}{2}\overline{A}I_{m}\left(  rq\right)  ,
\end{align*}
where the constants are $A$ and $\overline{A}$, and the modified Bessel
functions $I_{m}\left(  rq\right)  $ are regular at $x=0$. Within the walls of
the tube, for the radial parts one finds%
\begin{align*}
A_{r}  &  =\frac{1}{2}(-B\varepsilon Y_{m}^{\prime}\left(  rp\right)
+\overline{B}\varepsilon\frac{m}{rp}Y_{m}\left(  rp\right)  -C\varepsilon
J_{m}^{\prime}\left(  rp\right)  +\overline{C}\varepsilon\frac{m}{rp}%
J_{m}\left(  rp\right)  )\\
A_{\theta}  &  =\frac{1}{2}(-\overline{B}Y_{m}^{\prime}\left(  rp\right)
+B\frac{m}{rp}Y_{m}\left(  rp\right)  -\overline{C}J_{m}^{\prime}\left(
rp\right)  +C\frac{m}{rp}J_{m}\left(  rp\right)  )\\
A_{z}  &  =\frac{1}{2}\frac{p}{k}BY_{m}\left(  rp\right)  +\frac{1}{2}\frac
{p}{k}CJ_{m}\left(  rp\right) \\
B_{r}  &  =\frac{1}{2}(-\overline{B}kY_{m}^{\prime}\left(  rp\right)
+B\frac{n^{2}\omega^{2}}{kc^{2}}\frac{m}{rp}Y_{m}\left(  rp\right)
-\overline{C}kJ_{m}^{\prime}\left(  rp\right)  +C\frac{n^{2}\omega^{2}}%
{kc^{2}}\frac{m}{rp}J_{m}\left(  rp\right)  )\\
B_{\theta}  &  =\frac{1}{2}(\overline{B}k\frac{1}{\overline{\mu}}\frac{m}%
{rp}Y_{m}\left(  rp\right)  -\frac{1}{\mu}\frac{n^{2}\omega^{2}}{kc^{2}}%
BY_{m}^{\prime}\left(  rp\right)  +\overline{C}k\frac{1}{\overline{\mu}}%
\frac{m}{rp}J_{m}\left(  rp\right)  -\frac{1}{\overline{\mu}}\frac{n^{2}%
\omega^{2}}{kc^{2}}CJ_{m}^{\prime}\left(  rp\right)  )\\
\frac{1}{\overline{\mu}}B_{z}  &  =\frac{1}{\overline{\mu}}\frac{1}%
{2}p\overline{B}Y_{m}\left(  rp\right)  +\frac{1}{\overline{\mu}}\frac{1}%
{2}p\overline{C}J_{m}\left(  rp\right)  ,
\end{align*}
where $J_{m}\left(  x\right)  $ and $Y_{m}\left(  x\right)  $ are the Bessel
functions, the corresponding coefficients are $C$, $\overline{C}$, $B$, and
$\overline{B}$, while $p=\sqrt{\left(  n\frac{\omega}{c}\right)  ^{2}-k^{2}}$.
Here $n=\sqrt{\overline{\mu}\varepsilon}$ is the refraction index, where
$\overline{\mu}$ and $\varepsilon$ are the magnetic and the dielectric linear
susceptibilities, respectively.

Conditions of the tangential fields continuity at the inner $R_{inn}=R_{1}$
and the outer $R_{out}=R_{2}$ radii of the tube can be written as a product of
a vector by matrix%
\begin{align*}
&  \left(
\begin{array}
[c]{cccccccc}%
1 & 0 & \frac{p}{q}Y_{m}\left(  R_{1}p\right)  & 0 & \frac{p}{q}J_{m}\left(
R_{1}p\right)  & 0 & 0 & 0\\
0 & 0 & \frac{p}{q}Y_{m}\left(  R_{2}p\right)  & 0 & \frac{p}{q}J_{m}\left(
R_{2}p\right)  & 0 & 1 & 0\\
0 & 1 & 0 & \frac{1}{\mu}\frac{p}{q}Y_{m}\left(  R_{1}p\right)  & 0 & \frac
{1}{\mu}\frac{p}{q}J_{m}\left(  R_{1}p\right)  & 0 & 0\\
0 & 0 & 0 & \frac{1}{\mu}\frac{p}{q}Y_{m}\left(  R_{2}p\right)  & 0 & \frac
{1}{\mu}\frac{p}{q}J_{m}\left(  R_{2}p\right)  & 0 & 1\\
-\frac{m}{R_{1}q} & \frac{I_{m}^{\prime}\left(  R_{1}q\right)  }{I_{m}\left(
R_{1}q\right)  } & \frac{m}{R_{1}p}Y_{m}\left(  R_{1}p\right)  &
-Y_{m}^{\prime}\left(  R_{1}p\right)  & \frac{m}{R_{1}p}J_{m}\left(
R_{1}p\right)  & -J_{m}^{\prime}\left(  R_{1}p\right)  & 0 & 0\\
0 & 0 & \frac{m}{R_{2}p}Y_{m}\left(  R_{2}p\right)  & -Y_{m}^{\prime}\left(
R_{2}p\right)  & \frac{m}{R_{2}p}J_{m}\left(  R_{2}p\right)  & -J_{m}^{\prime
}\left(  R_{2}p\right)  & -\frac{m}{R_{2}q} & \frac{K_{m}^{\prime}\left(
R_{2}q\right)  }{K_{m}\left(  R_{2}q\right)  }\\
\frac{\omega^{2}}{kc^{2}}\frac{I_{m}^{\prime}\left(  R_{1}q\right)  }%
{I_{m}\left(  R_{1}q\right)  } & -\frac{km}{R_{1}q} & -\frac{1}{\mu}%
\frac{n^{2}\omega^{2}}{kc^{2}}Y_{m}^{\prime}\left(  R_{1}p\right)  & \frac
{1}{\mu}\frac{km}{R_{1}p}Y_{m}\left(  R_{1}p\right)  & -\frac{1}{\mu}%
\frac{n^{2}\omega^{2}}{kc^{2}}J_{m}^{\prime}\left(  R_{1}p\right)  & \frac
{1}{\mu}\frac{km}{R_{1}p}J_{m}\left(  R_{1}p\right)  & 0 & 0\\
0 & 0 & -\frac{1}{\mu}\frac{n^{2}\omega^{2}}{kc^{2}}Y_{m}^{\prime}\left(
R_{2}p\right)  & k\frac{1}{\mu}\frac{m}{R_{2}p}Y_{m}\left(  R_{2}p\right)  &
-\frac{1}{\mu}\frac{n^{2}\omega^{2}}{kc^{2}}J_{m}^{\prime}\left(
R_{2}p\right)  & k\frac{1}{\mu}\frac{m}{R_{2}p}J_{m}\left(  R_{2}p\right)  &
\frac{\omega^{2}}{kc^{2}}\frac{K_{m}^{\prime}\left(  R_{2}q\right)  }%
{K_{m}\left(  R_{2}q\right)  } & -\frac{km}{R_{2}q}%
\end{array}
\right) \\
&  \times\left(
\begin{array}
[c]{c}%
AI_{m}\left(  R_{1}q\right) \\
\overline{A}I_{m}\left(  R_{1}q\right) \\
B\\
\overline{B}\\
C\\
\overline{C}\\
DK_{m}\left(  rq\right) \\
\overline{D}K_{m}\left(  R_{2}q\right)
\end{array}
\right)  ,
\end{align*}
which should give zero vector for nonzero $\left(  A,\overline{A},\ldots
D,\overline{D}\right)  $. This implies, that the determinant of the matrix
above equals zero, and the vector multiplying this matrix is an eigenvector
corresponding to zero eigenvalue. This vector will give the coefficients
$A,\ldots,\overline{D}$ and thereby the fields distribution, corresponding to
the value of the wavevector $k\left(  \omega\right)  $ making the determinant
equal to zero.

In the following figure
\begin{center}
\includegraphics[
height=2.4144in,
width=3.6624in
]%
{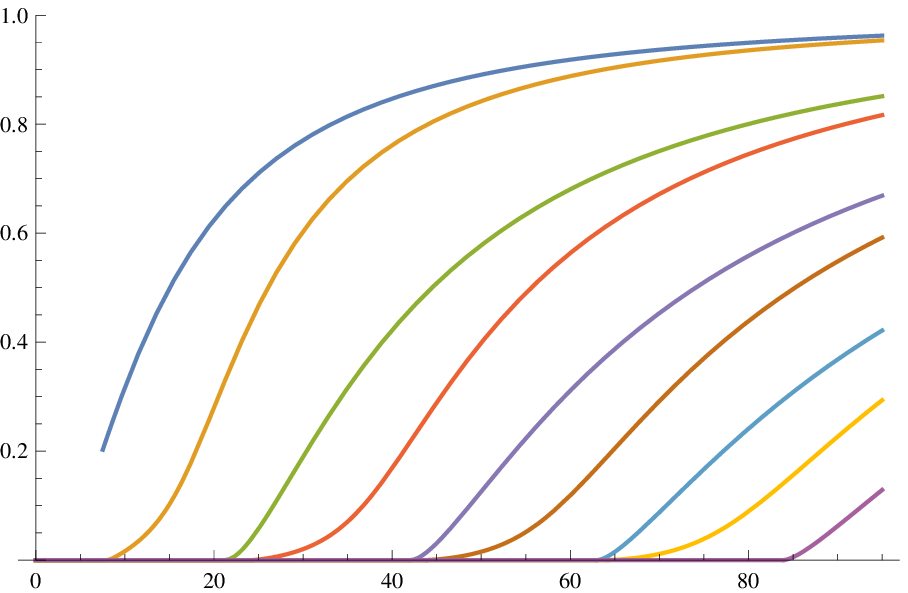}%
\\
Phase velocities $n=2.26$, $R_{2}/R_{1}=1.15$. $R_{1}=3\mu$, $\lambda_{1}%
=1\mu$, $\lambda_{2}=0.9992\mu$ On the abscise axis -- scaled frequency
$w=R_{inn}\frac{\omega}{c}\sqrt{n^{2}-1}$, on the ordinate axis scaled and
shifted phase velocity $b=\frac{\frac{kc}{\omega}-1}{n-1}$ .
\label{PY}%
\end{center}
one sees dependences $k\left(  \omega\right)  $ that have been found
numerically for the double degenerate modes corresponding $m=\pm1$. Each of
the mode can carry a polarized photon, such that the quantum information can
be encoded in the photon polarization. In the following figure
\begin{center}
\includegraphics[
height=2.3842in,
width=3.6624in
]%
{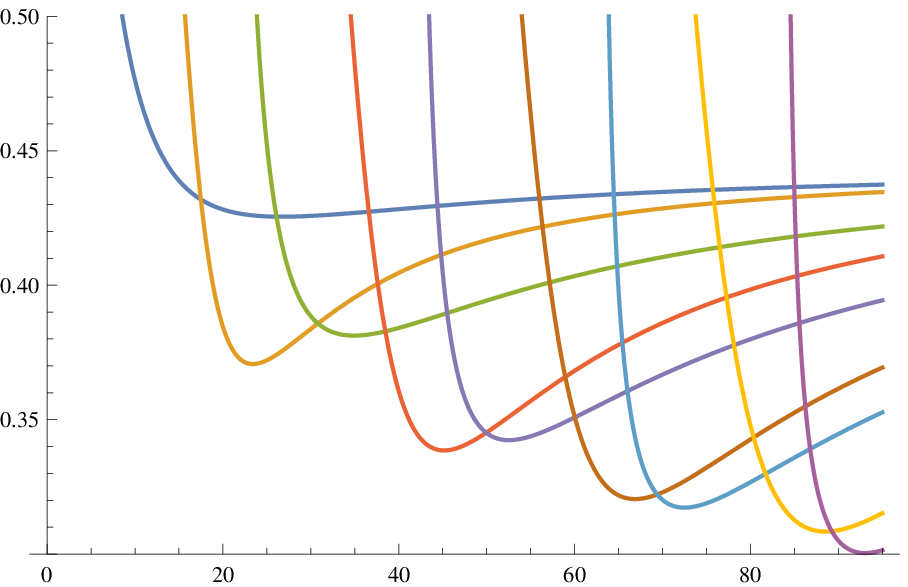}%
\\
Group velocities $n=2.26$. On the abscise axis -- scaled frequency
$w=R_{inn}\frac{\omega}{c}\sqrt{n^{2}-1}$, on the ordinate axis scaled group
velocity $v/c$ .
\label{GV}%
\end{center}
one sees the corresponding group velocities and the frequencies where the
group velocities of different modes coincide.

Field distribution for the components of vector potential corresponding to a
point\ close to the point of the group velocity coincidence of the first and
the second modes given by the corresponding coefficients are shown in the
following figure%
\begin{center}
\includegraphics[
height=2.8374in,
width=4.4464in
]%
{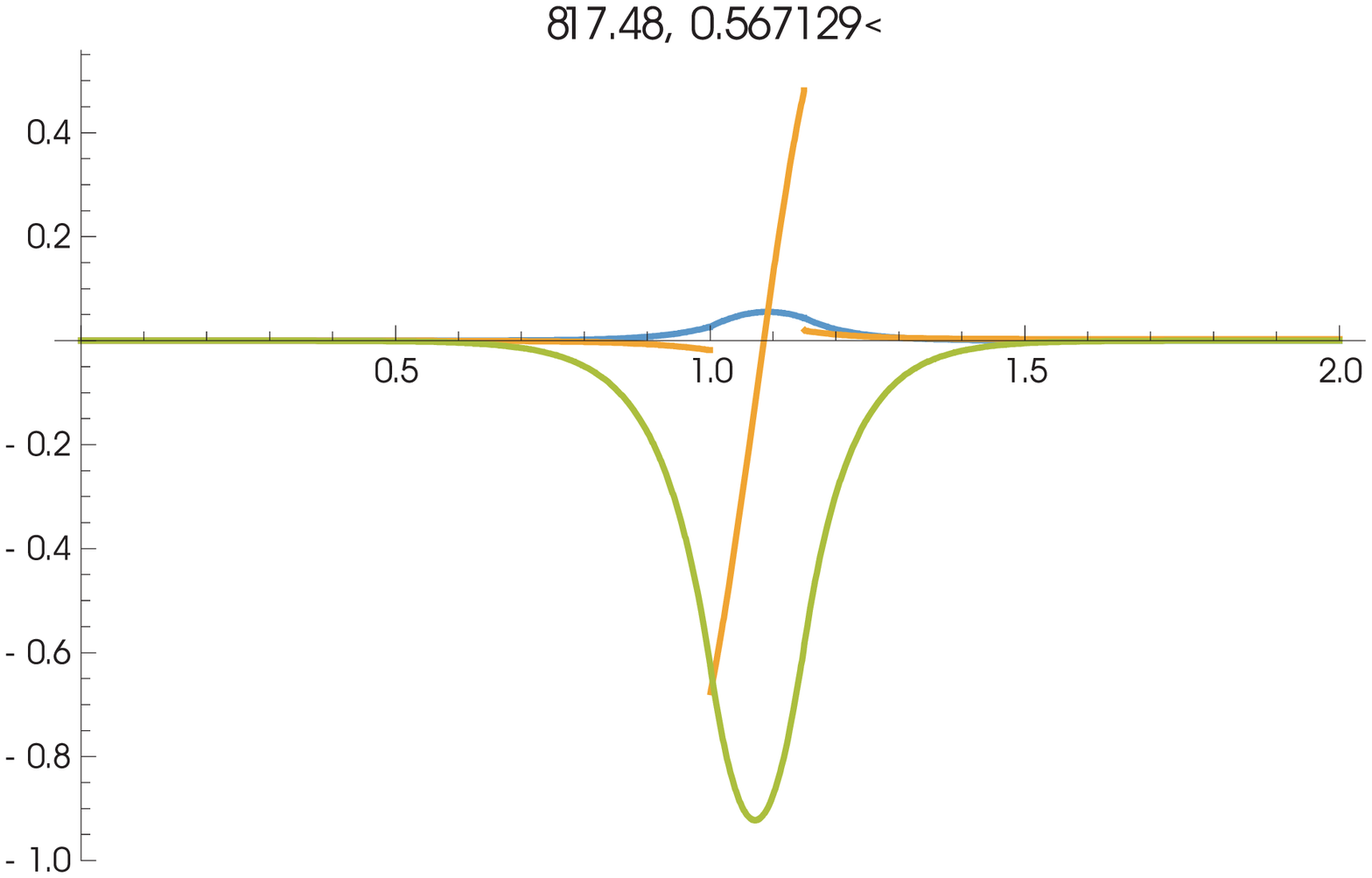}%
\\
Vector potential components for the first mode: the radial -blue, the azimutal
component -brown, and the longitudinal component - green. Numbers at the plot
are the coordinates $w$ and $b$ of the Figure for the phase velocity
\label{Mode1}%
\end{center}
and the figure%
\begin{center}
\includegraphics[
height=2.7541in,
width=4.2797in
]%
{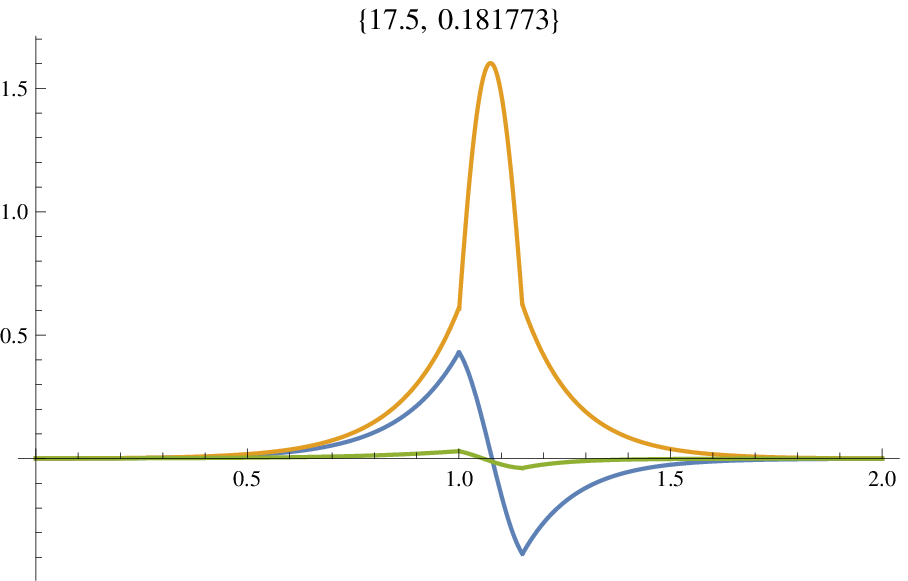}%
\\
Vector potential components for the second mode: the radial -blue, the
azimutal component -brown, and the longitudinal component - green. Numbers at
the plot are the coordinates $w$ and $b$ of the Figure for the phase velocity
\label{Mode2}%
\end{center}
for the first and the second modes, respectively. The mode fields are
normalized by the requirement $\int\left(  \overrightarrow{u}_{k}^{\ast
}\left(  R\right)  \cdot\overrightarrow{u}_{k}\left(  R\right)  \right)
RdR=1$

One can equally find the scalar product of the vector potential of the first
mode by that of the second one. The following figure%
\begin{center}
\includegraphics[
height=3.0911in,
width=4.9068in
]%
{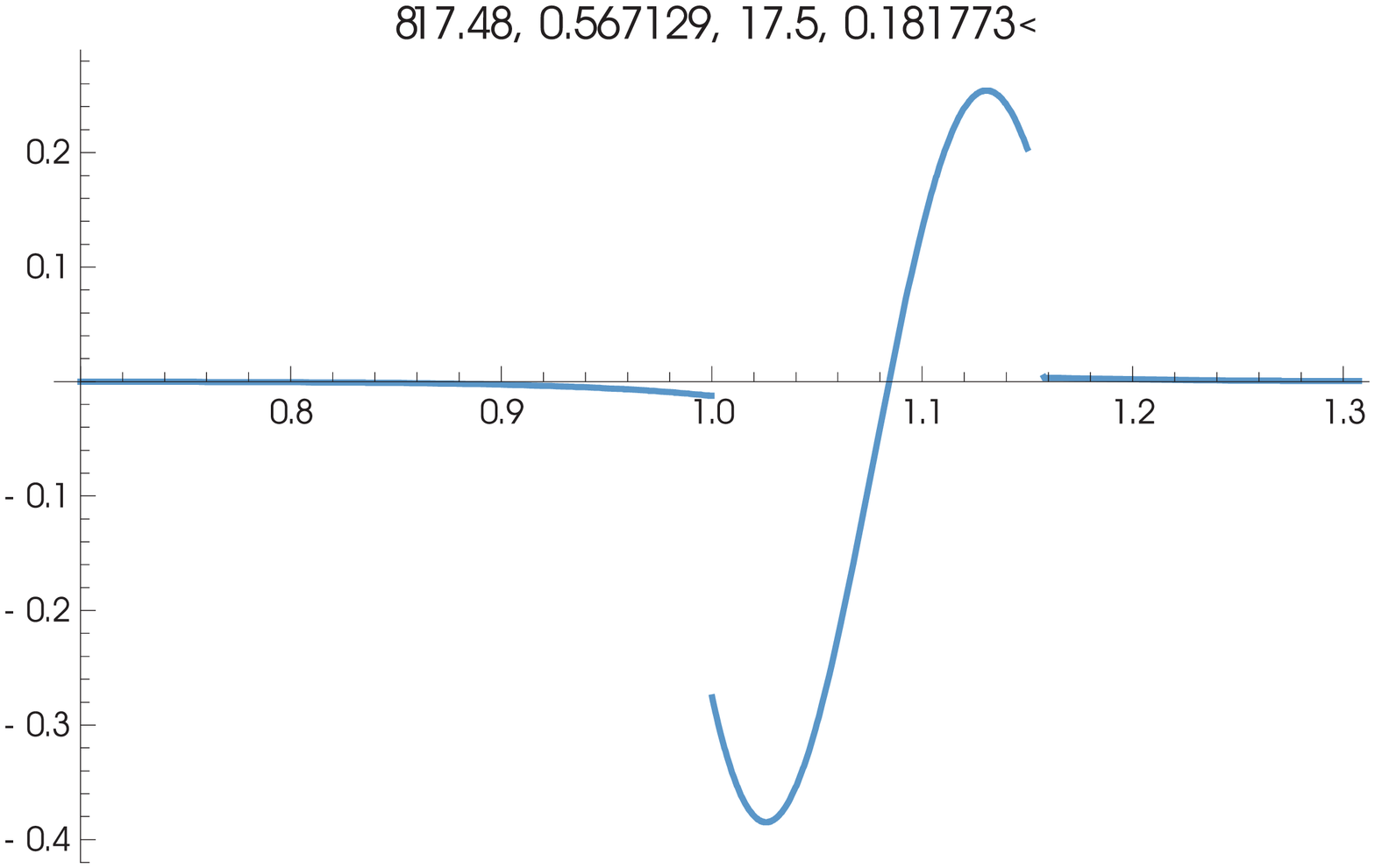}%
\\
Scalar product of mode vector potentials as a function of the radius.\ The
maximum product $-0.4$ corresponds to $R=1.05\ R_{inn}$. The frequency
difference $\delta\omega$ corresponds to the wavelength difference of
$0,96mm$. Numbers at the plot are the coordinates $w$ and $b$ of the Figure
for the phase velocity. Normalization $\int\left(  \overrightarrow{u}%
_{k}^{\ast}\left(  R\right)  \cdot\overrightarrow{u}_{k^{\prime}}\left(
R\right)  \right)  RdR=1$ is done in the dimensionless unities $R_{inn}=1$.
\label{OL}%
\end{center}
shows the dependence of the scalar product on the radius. Note that the
integral $\int\left(  \overrightarrow{u}_{k}^{\ast}\left(  R\right)
\cdot\overrightarrow{u}_{k^{\prime}}\left(  R\right)  \right)  RdR$ of the
scalar product vanishes for equal frequencies $\omega=\omega^{\prime}$. One
sees that this is almost the case at the lst figure

\section{Appendix E\label{Appendix E}}

One considers the action integrand%
\begin{equation}
Lg=\ln\left[  \det\left(
\begin{array}
[c]{cccc}%
\delta\omega+\widetilde{\omega}-\text{$\epsilon_{1}$} & \alpha^{\ast
}\overrightarrow{u}_{k}^{\ast}\overrightarrow{u}_{k^{\prime}} & -\Delta
\text{$_{1}$} & -\Delta\\
\alpha\overrightarrow{u}_{k^{\prime}}^{\ast}\overrightarrow{u}_{k} &
\widetilde{\omega}-\text{$\epsilon_{2}$} & -\Delta & -\Delta_{2}\\
-\Delta^{\ast}\text{$_{1}$} & -\text{$\Delta^{\ast}$} & \widetilde{\omega
}+\text{$\epsilon_{3}$} & -\alpha\overrightarrow{u}_{k^{\prime}}^{\ast
}\overrightarrow{u}_{k}\\
-\text{$\Delta^{\ast}$} & -\Delta_{2}^{\ast} & -\alpha^{\ast}\overrightarrow
{u}_{k}^{\ast}\overrightarrow{u}_{k^{\prime}} & \delta\omega+\widetilde
{\omega}+\text{$\epsilon_{4}$}%
\end{array}
\right)  \right]  .
\end{equation}
nd performs the Taylor expansion%
\begin{equation}
\left(
\begin{array}
[c]{ccc}%
\alpha^{\ast}\overrightarrow{u}_{k}^{\ast}\overrightarrow{u}_{k^{\prime}} &
\frac{\Delta_{1}}{\sqrt{2\pi dRl}} & \frac{\Delta_{2}^{\ast}}{\sqrt{2\pi dRl}}%
\end{array}
\right)  \left(
\begin{array}
[c]{ccc}%
\frac{\partial^{2}Lg}{\partial\alpha^{\ast}\overrightarrow{u}_{k}^{\ast
}\overrightarrow{u}_{k^{\prime}}\partial\alpha\overrightarrow{u}_{k^{\prime}%
}^{\ast}\overrightarrow{u}_{k}} & \frac{\partial^{2}Lg}{\partial\Delta
_{1}^{\ast}\partial\alpha\overrightarrow{u}_{k^{\prime}}^{\ast}\overrightarrow
{u}_{k}} & \frac{\partial^{2}Lg}{\partial\Delta_{2}\partial\alpha
\overrightarrow{u}_{k^{\prime}}^{\ast}\overrightarrow{u}_{k}}\\
\frac{\partial^{2}Lg}{\partial\alpha^{\ast}\overrightarrow{u}_{k}^{\ast
}\overrightarrow{u}_{k^{\prime}}\partial\Delta\text{$_{1}$}} & \frac
{\partial^{2}Lg}{\partial\Delta_{1}^{\ast}\partial\Delta\text{$_{1}$}} &
\frac{\partial^{2}Lg}{\partial\Delta_{2}\partial\Delta\text{$_{1}$}}\\
\frac{\partial^{2}Lg}{\partial\alpha^{\ast}\overrightarrow{u}_{k}^{\ast
}\overrightarrow{u}_{k^{\prime}}\partial\Delta^{\ast}\text{$_{2}$}} &
\frac{\partial^{2}Lg}{\partial\Delta_{1}^{\ast}\partial\Delta^{\ast
}\text{$_{2}$}} & \frac{\partial^{2}Lg}{\partial\Delta_{2}\partial\Delta
^{\ast}\text{$_{2}$}}%
\end{array}
\right)  \left(
\begin{array}
[c]{c}%
\alpha\overrightarrow{u}_{k^{\prime}}^{\ast}\overrightarrow{u}_{k}\\
\frac{\Delta_{1}^{\ast}}{\sqrt{2\pi dRl}}\\
\frac{\Delta_{2}}{\sqrt{2\pi dRl}}%
\end{array}
\right)  . \label{EQTeylor}%
\end{equation}
Explicit form of the matrix reads
\begin{align}
\frac{\partial^{2}Lg}{\partial\ldots\partial\ldots}  &  =\left(
\begin{array}
[c]{ccc}%
\frac{\partial^{2}Lg}{\partial\alpha^{\ast}\overrightarrow{u}_{k}^{\ast
}\overrightarrow{u}_{k^{\prime}}\partial\alpha\overrightarrow{u}_{k^{\prime}%
}^{\ast}\overrightarrow{u}_{k}} & \frac{\partial^{2}Lg}{\partial\Delta
_{1}^{\ast}\partial\alpha\overrightarrow{u}_{k^{\prime}}^{\ast}\overrightarrow
{u}_{k}} & \frac{\partial^{2}Lg}{\partial\Delta_{2}\partial\alpha
\overrightarrow{u}_{k^{\prime}}^{\ast}\overrightarrow{u}_{k}}\\
\frac{\partial^{2}Lg}{\partial\alpha^{\ast}\overrightarrow{u}_{k}^{\ast
}\overrightarrow{u}_{k^{\prime}}\partial\Delta\text{$_{1}$}} & \frac
{\partial^{2}Lg}{\partial\Delta_{1}^{\ast}\partial\Delta\text{$_{1}$}} &
\frac{\partial^{2}Lg}{\partial\Delta_{2}\partial\Delta\text{$_{1}$}}\\
\frac{\partial^{2}Lg}{\partial\alpha^{\ast}\overrightarrow{u}_{k}^{\ast
}\overrightarrow{u}_{k^{\prime}}\partial\Delta^{\ast}\text{$_{2}$}} &
\frac{\partial^{2}Lg}{\partial\Delta_{1}^{\ast}\partial\Delta^{\ast
}\text{$_{2}$}} & \frac{\partial^{2}Lg}{\partial\Delta_{2}\partial\Delta
^{\ast}\text{$_{2}$}}%
\end{array}
\right) \label{EQTaylor1}\\
&  \equiv\left(
\begin{array}
[c]{ccc}%
Lg_{\alpha,\alpha} & \frac{\Delta}{\left\vert \Delta\right\vert }%
Lg_{\Delta_{1},\alpha} & \frac{\Delta^{\ast}}{\left\vert \Delta\right\vert
}Lg_{\Delta_{2},\alpha}\\
\frac{\Delta^{\ast}}{\left\vert \Delta\right\vert }Lg_{\Delta_{1},\alpha} &
Lg_{\Delta_{1},\Delta_{1}} & \frac{\Delta^{\ast2}}{\left\vert \Delta
\right\vert ^{2}}Lg_{\Delta_{1},\Delta_{2}}\\
\frac{\Delta}{\left\vert \Delta\right\vert }Lg_{\Delta_{2},\alpha} &
\frac{\Delta^{2}}{\left\vert \Delta\right\vert ^{2}}Lg_{\Delta_{1},\Delta_{2}}
& Lg_{\Delta_{2},\Delta_{2}}%
\end{array}
\right) \nonumber
\end{align}
where%
\begin{align}
Lg_{\alpha,\alpha}  &  =\frac{-2\widetilde{\omega}^{2}+(\text{$\epsilon_{1}$%
}+\text{$\epsilon_{2}$}-\text{$\epsilon_{3}$}-\text{$\epsilon_{4}$}%
)\widetilde{\omega}+2\left\vert \Delta\right\vert ^{2}-\text{$\epsilon
_{1}\epsilon_{2}$}-\text{$\epsilon_{3}\epsilon_{4}$}+\delta\omega
(\text{$\epsilon_{2}$}-\text{$\epsilon_{3}$}-2\widetilde{\omega})}{(\left\vert
\Delta\right\vert ^{2}+(\text{$\epsilon_{2}$}-\widetilde{\omega}%
)(\text{$\epsilon_{3}$}+\widetilde{\omega}))(\left\vert \Delta\right\vert
^{2}-(\delta\omega-\text{$\epsilon_{1}$}+\widetilde{\omega})(\delta
\omega+\epsilon_{4}+\widetilde{\omega}))}\nonumber\\
Lg_{\Delta_{1},\Delta_{1}}  &  =\frac{(\text{$\epsilon_{2}$}-\omega
)(\delta\omega+\text{$\epsilon_{4}$}+\widetilde{\omega})}{(\left\vert
\Delta\right\vert ^{2}+(\text{$\epsilon_{2}$}-\widetilde{\omega}%
)(\text{$\epsilon_{3}$}+\widetilde{\omega}))(\left\vert \Delta\right\vert
^{2}-(\delta\omega-\text{$\epsilon_{1}$}+\widetilde{\omega})(\delta
\omega+\epsilon_{4}+\widetilde{\omega}))}\nonumber\\
Lg_{\Delta_{2},\Delta_{2}}  &  =\frac{-(\delta\omega-\text{$\epsilon_{1}$%
}+\omega)(\text{$\epsilon_{3}$}+\widetilde{\omega})}{(\left\vert
\Delta\right\vert ^{2}+(\text{$\epsilon_{2}$}-\widetilde{\omega}%
)(\text{$\epsilon_{3}$}+\widetilde{\omega}))(\left\vert \Delta\right\vert
^{2}-(\delta\omega-\text{$\epsilon_{1}$}+\widetilde{\omega})(\delta
\omega+\epsilon_{4}+\widetilde{\omega}))}\nonumber\\
Lg_{\Delta_{1},\alpha}  &  =\frac{\left\vert \Delta\right\vert (\delta
\omega+\text{$\epsilon_{2}$}+\text{$\epsilon_{4}$})}{(\left\vert
\Delta\right\vert ^{2}+(\text{$\epsilon_{2}$}-\widetilde{\omega}%
)(\text{$\epsilon_{3}$}+\widetilde{\omega}))(\left\vert \Delta\right\vert
^{2}-(\delta\omega-\text{$\epsilon_{1}$}+\widetilde{\omega})(\delta
\omega+\epsilon_{4}+\widetilde{\omega}))}\nonumber\\
Lg_{\Delta_{2},\alpha}  &  =\frac{\left\vert \text{$\Delta$}\right\vert
(-\delta\omega+\text{$\epsilon_{1}$}+\text{$\epsilon_{3}$})}{(\left\vert
\Delta\right\vert ^{2}+(\text{$\epsilon_{2}$}-\widetilde{\omega}%
)(\text{$\epsilon_{3}$}+\widetilde{\omega}))(\left\vert \Delta\right\vert
^{2}-(\delta\omega-\text{$\epsilon_{1}$}+\widetilde{\omega})(\delta
\omega+\epsilon_{4}+\widetilde{\omega}))}\nonumber\\
Lg_{\Delta_{1},\Delta_{2}}  &  =\frac{-\left\vert \Delta\right\vert ^{2}%
}{(\left\vert \Delta\right\vert ^{2}+(\text{$\epsilon_{2}$}-\widetilde{\omega
})(\text{$\epsilon_{3}$}+\widetilde{\omega}))(\left\vert \Delta\right\vert
^{2}-(\delta\omega-\text{$\epsilon_{1}$}+\widetilde{\omega})(\delta
\omega+\epsilon_{4}+\widetilde{\omega}))} \label{LG}%
\end{align}
One finds roots of the denominator and cast it in the form of product%
\begin{align*}
&  \left(  \widetilde{\omega}-\sqrt{\left\vert \Delta\right\vert ^{2}+\left(
\frac{\text{$\epsilon_{2}+\epsilon_{3}$}}{2}\right)  ^{2}}\text{$+$}%
\frac{\text{$\epsilon_{3}$}-\text{$\epsilon_{2}$}}{2}-io\right)  \left(
\widetilde{\omega}+\sqrt{\left\vert \Delta\right\vert ^{2}+\left(
\frac{\text{$\epsilon_{2}+\epsilon_{3}$}}{2}\right)  ^{2}}\text{$+$}%
\frac{\text{$\epsilon_{3}$}-\text{$\epsilon_{2}$}}{2}+io\right) \\
&  \left(  \widetilde{\omega}-\sqrt{\left\vert \Delta\right\vert ^{2}+\left(
\frac{\text{$\epsilon_{1}+\epsilon_{4}$}}{2}\right)  ^{2}}+\delta
\omega\text{$+$}\frac{\text{$\epsilon_{4}$}-\text{$\epsilon_{1}$}}%
{2}-io\right)  \left(  \widetilde{\omega}+\sqrt{\left\vert \Delta\right\vert
^{2}+\left(  \frac{\text{$\epsilon_{1}+\epsilon_{4}$}}{2}\right)  ^{2}}%
+\delta\omega\text{$+$}\frac{\text{$\epsilon_{4}$}-\text{$\epsilon_{1}$}}%
{2}+io\right)  ,
\end{align*}
allowing for the correct rule of the poles circumvention, which implies that
the virtual transition occur from the occupied states of the pairs below the
gap to the empty states of the pairs above the gap.

Putting apart the terms of Eq.(\ref{LG}) in such a way that the frequency
dependent factors in the denominator are grouped in pairs $\left(
\widetilde{\omega}-\sqrt{\left\vert \Delta\right\vert ^{2}+\left(
\frac{\text{$\epsilon_{2}+\epsilon_{3}$}}{2}\right)  ^{2}}\text{$+$}%
\frac{\text{$\epsilon_{3}$}-\text{$\epsilon_{2}$}}{2}-io\right)  $
$\times\left(  \widetilde{\omega}+\sqrt{\left\vert \Delta\right\vert
^{2}+\left(  \frac{\text{$\epsilon_{1}+\epsilon_{4}$}}{2}\right)  ^{2}}%
+\delta\omega\text{$+$}\frac{\text{$\epsilon_{4}$}-\text{$\epsilon_{1}$}}%
{2}+io\right)  $, and $\left(  \widetilde{\omega}+\sqrt{\left\vert
\Delta\right\vert ^{2}+\left(  \frac{\text{$\epsilon_{2}+\epsilon_{3}$}}%
{2}\right)  ^{2}}\text{$+$}\frac{\text{$\epsilon_{3}$}-\text{$\epsilon_{2}$}%
}{2}+io\right)  $ $\times$ $\left(  \widetilde{\omega}-\sqrt{\left\vert
\Delta\right\vert ^{2}+\left(  \frac{\text{$\epsilon_{1}+\epsilon_{4}$}}%
{2}\right)  ^{2}}+\delta\omega\text{$+$}\frac{\text{$\epsilon_{4}$%
}-\text{$\epsilon_{1}$}}{2}-io\right)  $, and integrating over $d\widetilde
{\omega}$ , yields the matrix%
\begin{align}
\widehat{\widetilde{M}}  &  =\int d\widetilde{\omega}\frac{\partial^{2}%
}{\partial\ldots\partial\ldots}Lg\label{EQM}\\
&  \equiv\left(
\begin{array}
[c]{ccc}%
\widetilde{M}_{\alpha,\alpha} & \frac{\Delta}{\left\vert \Delta\right\vert
}\widetilde{M}_{\Delta_{1},\alpha} & \frac{\Delta^{\ast}}{\left\vert
\Delta\right\vert }\widetilde{M}_{\Delta_{2},\alpha}\\
\frac{\Delta^{\ast}}{\left\vert \Delta\right\vert }\widetilde{M}_{\Delta
_{1},\alpha} & \widetilde{M}_{\Delta_{1},\Delta_{1}} & \frac{\Delta^{\ast2}%
}{\left\vert \Delta\right\vert ^{2}}\widetilde{M}_{\Delta_{1},\Delta_{2}}\\
\frac{\Delta}{\left\vert \Delta\right\vert }\widetilde{M}_{\Delta_{2},\alpha}
& \frac{\Delta^{2}}{\left\vert \Delta\right\vert ^{2}}\widetilde{M}%
_{\Delta_{1},\Delta_{2}} & \widetilde{M}_{\Delta_{2},\Delta_{2}}%
\end{array}
\right) \nonumber
\end{align}
with the matrix elements%
\begin{equation}%
\begin{array}
[c]{c}%
\widetilde{M}_{\alpha,\alpha}=-\frac{2i\pi\left(  4\left\vert \Delta
\right\vert ^{2}-\left(  \epsilon_{2}+\epsilon_{3}\right)  \left(
\epsilon_{1}+\epsilon_{4}\right)  +\eta_{1}\eta_{2}\right)  }{\eta_{1}\eta
_{2}\left(  \delta\epsilon-2\delta\omega+\eta_{1}+\eta_{2}\right)  }%
-\frac{2i\pi\left(  4\left\vert \Delta\right\vert ^{2}-\left(  \epsilon
_{2}+\epsilon_{3}\right)  \left(  \epsilon_{1}+\epsilon_{4}\right)  +\eta
_{1}\eta_{2}\right)  }{\eta_{1}\eta_{2}\left(  -\delta\epsilon+2\delta
\omega+\eta_{1}+\eta_{2}\right)  },\\
\widetilde{M}_{\Delta_{1},\Delta_{1}}=-\frac{i\pi\left(  -\epsilon
_{2}-\epsilon_{3}+\eta_{1}\right)  \left(  -\epsilon_{1}-\epsilon_{4}+\eta
_{2}\right)  }{\eta_{1}\eta_{2}\left(  -\delta\epsilon+2\delta\omega+\eta
_{1}+\eta_{2}\right)  }-\frac{i\pi\left(  \epsilon_{2}+\epsilon_{3}+\eta
_{1}\right)  \left(  \epsilon_{1}+\epsilon_{4}+\eta_{2}\right)  }{\eta_{1}%
\eta_{2}\left(  \delta\epsilon-2\delta\omega+\eta_{1}+\eta_{2}\right)  },\\
\widetilde{M}_{\Delta_{2},\Delta_{2}}=-\frac{i\pi\left(  -\epsilon
_{2}-\epsilon_{3}+\eta_{1}\right)  \left(  -\epsilon_{1}-\epsilon_{4}+\eta
_{2}\right)  }{\eta_{1}\eta_{2}\left(  \delta\epsilon-2\delta\omega+\eta
_{1}+\eta_{2}\right)  }-\frac{i\pi\left(  \epsilon_{2}+\epsilon_{3}+\eta
_{1}\right)  \left(  \epsilon_{1}+\epsilon_{4}+\eta_{2}\right)  }{\eta_{1}%
\eta_{2}\left(  2\delta\omega-\delta\epsilon+\eta_{1}+\eta_{2}\right)  },\\
\widetilde{M}_{\Delta_{1},\alpha}=\frac{2i\pi\Delta\left(  -\epsilon_{s}%
+\eta_{1}+\eta_{2}\right)  }{\eta_{1}\eta_{2}\left(  -\delta\epsilon
+2\delta\omega+\eta_{1}+\eta_{2}\right)  }-\frac{2i\pi\Delta\left(
\epsilon_{s}+\eta_{1}+\eta_{2}\right)  }{\eta_{1}\eta_{2}\left(
\delta\epsilon-2\delta\omega+\eta_{1}+\eta_{2}\right)  },\\
\widetilde{M}_{\Delta_{2},\alpha}=\frac{2i\pi\left\vert \Delta\right\vert
\left(  -\epsilon_{s}+\eta_{1}+\eta_{2}\right)  }{\eta_{1}\eta_{2}\left(
\delta\epsilon-2\delta\omega+\eta_{1}+\eta_{2}\right)  }-\frac{2i\pi\left\vert
\Delta\right\vert \left(  \epsilon_{s}+\eta_{1}+\eta_{2}\right)  }{\eta
_{1}\eta_{2}\left(  -\delta\epsilon+2\delta\omega+\eta_{1}+\eta_{2}\right)
},\\
\widetilde{M}_{\Delta_{1},\Delta_{2}}=\frac{4i\pi\left\vert \Delta\right\vert
^{2}}{\eta_{1}\eta_{2}\left(  \delta\epsilon-2\delta\omega+\eta_{1}+\eta
_{2}\right)  }+\frac{4i\pi\left\vert \Delta\right\vert ^{2}}{\eta_{1}\eta
_{2}\left(  -\delta\epsilon+2\delta\omega+\eta_{1}+\eta_{2}\right)  },
\end{array}
\label{EQ36}%
\end{equation}
where the combinations $\eta_{2}=\sqrt{4\left\vert \Delta\right\vert
^{2}+\left(  \epsilon_{1}+\epsilon_{4}\right)  {}^{2}}$ and $\eta_{1}%
=\sqrt{4\left\vert \Delta\right\vert ^{2}+\left(  \epsilon_{2}+\epsilon
_{3}\right)  {}^{2}}$ can be interpreted as energies of the initial and the
virtual final states of the Cooper pair, respectively. The notations
$\epsilon_{s}=\epsilon_{1}+\epsilon_{2}+\epsilon_{3}+\epsilon_{4}$, and
$\delta\epsilon=\epsilon_{1}-\epsilon_{2}+\epsilon_{3}-\epsilon_{4}$ are
introduced for shortness.

Tracing in Eq.(\ref{EQ33}) means that the expressions Eq.(\ref{EQ36})
containing energies $\epsilon_{i}$ should be integrated over the momenta
$p_{r}$ and $\widetilde{k}$ and summed over the angular momentum
$\widetilde{L}$. Performing this integration for the electron energy%
\begin{equation}
\text{$\epsilon_{f}$}(\widetilde{L},\widetilde{k})=\frac{p_{r}^{2}}{2}%
+\frac{\widetilde{k}^{2}}{2}+\frac{\left(  \widetilde{L}-\overline{L}\right)
^{2}}{2R^{2}}-\mu, \label{EQ37}%
\end{equation}
one can take into account that $\delta\epsilon=2\delta L(\Lambda-\overline
{L})/R^{2}$ and employ different integration variables: $\xi$, $\varsigma$,
and $L$, such that $\epsilon_{2}+\epsilon_{3}=2\left\vert \Delta\right\vert
\sinh\xi$, $\epsilon_{1}+\epsilon_{4}=2\left\vert \Delta\right\vert
\sinh\varsigma$, and $\widetilde{L}=L+\Lambda-\delta L/2+\frac{\left\vert
\Delta\right\vert \delta L\left(  \sinh\varsigma-\sinh\xi\right)  }%
{\delta\text{$k^{2}+\delta L^{2}/R^{2}$}}$, with the Jacobian%
\begin{equation}
J=\frac{4R^{2}\left\vert \text{$\Delta$}\right\vert ^{2}\cosh\zeta\cosh\xi
}{\left(  2\pi\right)  ^{3}\sqrt{\Gamma-4L^{2}\left(  \text{$\delta L$}%
^{2}+R^{2}\text{$\delta k$}^{2}\right)  }} \label{EQJacobi}%
\end{equation}
where%
\[
\Gamma=R^{4}\text{$\delta k$}^{2}\left[  4\left\vert \text{$\Delta$%
}\right\vert \left(  \sinh\zeta+\sinh\xi-\frac{\left\vert \text{$\Delta$%
}\right\vert }{\frac{\text{$\delta L$}^{2}}{R^{2}}+\text{$\delta k$}^{2}%
}\left(  \sinh\xi-\sinh\zeta\right)  ^{2}\right)  +\left(  8\mu-\text{$\delta
k$}^{2}-\frac{\text{$\delta L$}^{2}+4(\overline{L}-\Lambda)^{2}}{R^{2}%
}\right)  \right]  .
\]
It includes the phase volume factor $\left(  2\pi\right)  ^{3}$ and an
additional factor $2$ allowing for the other brunch corresponding to the
negative momenta $p_{r}$.

The variable $L$ enters only the Jacobian Eq.(\ref{EQJacobi}), and therefore
the latter can be integrated over this variable within the domain where the
square root is positive, thus yielding%
\begin{equation}
J_{L}\equiv\int\frac{dL}{R}\frac{4R^{2}\left\vert \text{$\Delta$}\right\vert
^{2}\cosh\zeta\cosh\xi}{\left(  2\pi\right)  ^{2}\sqrt{\Gamma-4L^{2}\left(
\text{$\delta L$}^{2}+R^{2}\text{$\delta k$}^{2}\right)  }}=\int\frac
{dX}{\sqrt{1-X^{2}}}\frac{2\left\vert \text{$\Delta$}\right\vert ^{2}%
\cosh\zeta\cosh\xi}{\left(  2\pi\right)  ^{3}\sqrt{\left(  \text{$\delta L$%
}^{2}/R^{2}+\text{$\delta k$}^{2}\right)  }}=\frac{\left\vert \text{$\Delta$%
}\right\vert ^{2}\cosh\zeta\cosh\xi}{\left(  2\pi\right)  ^{2}\sqrt
{\frac{\text{$\delta L$}^{2}}{R^{2}}+\text{$\delta k$}^{2}}}. \label{EqJac1}%
\end{equation}
The contribution differs from zero only if $\Gamma>0$, which determines the
integration domain $D\left[  J\right]  $ over the variables $\zeta$ and $\xi$
in Eq.(\ref{EQ45}). One thus arrives at%
\begin{equation}
\sinh\zeta+\sinh\xi-\frac{\left\vert \text{$\Delta$}\right\vert }%
{\frac{\text{$\delta L$}^{2}}{R^{2}}+\text{$\delta k$}^{2}}\left(  \sinh
\xi-\sinh\zeta\right)  ^{2}+\frac{2\mu}{\left\vert \text{$\Delta$}\right\vert
}>0, \label{EqBorder}%
\end{equation}
where the small term $-\delta k$$^{2}-\frac{\text{$\delta L$}^{2}%
+4(\overline{L}-\Lambda)^{2}}{R^{2}}$ is ignored as compared to $\mu$.

For the variables $\zeta=A+B/2$ and \ $\xi=A-B/2$, one can explicitly find the
borders of the integration domain $D[J]$. In fact Eq.(\ref{EqBorder}) in these
variables reads%
\[
2\cosh\frac{B}{2}\sinh A-\frac{4\left\vert \text{$\Delta$}\right\vert }%
{\frac{\text{$\delta L$}^{2}}{R^{2}}+\text{$\delta k$}^{2}}\left(  1+\sinh
^{2}A\right)  \sinh^{2}\frac{B}{2}+\frac{2\mu}{\left\vert \text{$\Delta$%
}\right\vert }>0
\]
and determines borders for the variable $A$:%
\[
\frac{\cosh\frac{B}{2}-\sqrt{\cosh^{2}\frac{B}{2}+16\kappa^{2}\sinh^{2}%
\frac{B}{2}(\frac{\mu}{\left\vert \text{$\Delta$}\right\vert 2\kappa}%
-\sinh^{2}\frac{B}{2})}}{4\kappa\sinh^{2}\frac{B}{2}}<\sinh A<\frac{\cosh
\frac{B}{2}+\sqrt{\cosh^{2}\frac{B}{2}+16\kappa^{2}\sinh^{2}\frac{B}{2}%
(\frac{\mu}{\left\vert \text{$\Delta$}\right\vert 2\kappa}-\sinh^{2}\frac
{B}{2})}}{4\kappa\sinh^{2}\frac{B}{2}},
\]
and this condition implies real borders, that is%
\[
\cosh^{2}\frac{B}{2}+16\kappa^{2}\sinh^{2}\frac{B}{2}(\frac{\mu}{\left\vert
\text{$\Delta$}\right\vert 2\kappa}-\sinh^{2}\frac{B}{2})>0,
\]
where $\kappa=\frac{2\left\vert \text{$\Delta$}\right\vert }{\frac
{\text{$\delta L$}^{2}}{R^{2}}+\text{$\delta k$}^{2}}$. Since $\cosh^{2}%
\frac{B}{2}=1+\sinh^{2}\frac{B}{2}$, one finds%
\[
\frac{8\kappa\frac{\mu}{\left\vert \text{$\Delta$}\right\vert }+1-\sqrt
{64\kappa^{2}+\left(  8\kappa\frac{\mu}{\left\vert \text{$\Delta$}\right\vert
}+1\right)  ^{2}}}{32\kappa^{2}}<\sinh^{2}\frac{B}{2}<\frac{8\kappa\frac{\mu
}{\left\vert \text{$\Delta$}\right\vert }+1+\sqrt{64\kappa^{2}+\left(
8\kappa\frac{\mu}{\left\vert \text{$\Delta$}\right\vert }+1\right)  ^{2}}%
}{32\kappa^{2}}.
\]
The left part is negative and hence this inequality always holds, while the
right part yields the integration domain over $B$.

The matrix elements Eq.(\ref{EQ36}) now read%
\begin{equation}%
\begin{array}
[c]{c}%
\widetilde{M}_{\alpha,\alpha}=-\widetilde{J}\left(  \frac{2i\pi(\cosh
(\zeta-\xi)+1)}{-\Omega+\cosh\zeta+\cosh\xi}+\frac{2i\pi(\cosh(\zeta-\xi
)+1)}{\Omega+\cosh\zeta+\cosh\xi}\right) \\
\widetilde{M}_{\Delta_{1},\Delta_{1}}=\widetilde{M}_{\Delta_{2},\Delta_{2}%
}=-\widetilde{J}\left(  \frac{i\pi e^{-\zeta-\xi}}{\Omega+\cosh\zeta+\cosh\xi
}+\frac{i\pi e^{\zeta+\xi}}{-\Omega+\cosh\zeta+\cosh\xi}\right) \\
\widetilde{M}_{\Delta_{1},\alpha}=-\widetilde{M}_{\Delta_{2},\alpha
}=\widetilde{J}\left(  \frac{i\pi\left(  e^{-\zeta}+e^{-\xi}\right)  }%
{\Omega+\cosh\zeta+\cosh\xi}-\frac{i\pi\left(  e^{\zeta}+e^{\xi}\right)
}{-\Omega+\cosh\zeta+\cosh\xi}\right) \\
\widetilde{M}_{\Delta_{1},\Delta_{2}}=\widetilde{J}\left(  \frac{i\pi}%
{-\Omega+\cosh\zeta+\cosh\xi}+\frac{i\pi}{\Omega+\cosh\zeta+\cosh\xi}\right)
\end{array}
, \label{EQ38}%
\end{equation}
where $\Omega=\frac{\delta\epsilon-2\Delta\omega}{2\left\vert \text{$\Delta$%
}\right\vert }=\frac{\delta L(\Lambda-\overline{L})/R^{2}-\Delta\omega}%
{\Delta}$ stands for the scaled and shifted perturbation frequency. One
recognizes the structure of the integrals Eq.(\ref{EQ45}). The factor in front
of the matrix elements
\begin{equation}
\widetilde{J}=\frac{J_{L}}{2\left\vert \text{$\Delta$}\right\vert \cosh
\zeta\cosh\xi}=\frac{\left\vert \text{$\Delta$}\right\vert }{8\pi^{2}%
\sqrt{\frac{\text{$\delta L$}^{2}}{R^{2}}+\text{$\delta k$}^{2}}}
\label{EqJacFact}%
\end{equation}
originates from the Jacobian Eq.(\ref{EqJac1}) and incorporates the factor
$\left(  2\left\vert \text{$\Delta$}\right\vert \right)  ^{-1}$, which makes
the frequency $\Omega$ dimensionless. The phase of the order parameter does
not enter in the final result and can be set to zero.

Now calculate the integrals%
\begin{equation}%
\begin{array}
[c]{c}%
\int\limits_{D[J]}d\xi d\varsigma\frac{(\cosh(\zeta-\xi)+1)}{\Omega
+\cosh(\zeta)+\cosh(\xi)}=\int\limits_{-b}^{b}dB\int\limits_{a_{-}}^{a_{+}%
}dA\frac{\cosh B+1}{\Omega+2\cosh A\cosh\frac{B}{2}}\\
\int\limits_{D[J]}d\xi d\varsigma\frac{-\left(  e^{-\zeta-\xi}+1\right)
}{\Omega+\cosh(\zeta)+\cosh(\xi)}=-\int\limits_{-b}^{b}dB\int\limits_{a_{-}%
}^{a_{+}}dA\frac{e^{-2A}+1}{\Omega+2\cosh A\cosh\frac{B}{2}}\\
\int\limits_{D[J]}d\xi d\varsigma\frac{\left(  e^{-\zeta}+e^{-\xi}\right)
}{\Omega+\cosh\zeta+\cosh\xi}=2\int\limits_{-b}^{b}dB\int\limits_{a_{-}%
}^{a_{+}}dA\frac{e^{A}\cosh\frac{B}{2}}{\Omega+2\cosh A\cosh\frac{B}{2}}\\
a_{\pm}=\mathrm{arcsinh}\frac{\cosh\frac{B}{2}\pm\sqrt{\cosh^{2}\frac{B}%
{2}+16\kappa^{2}\sinh^{2}\frac{B}{2}(\frac{\mu}{\left\vert \text{$\Delta$%
}\right\vert 2\kappa}-\sinh^{2}\frac{B}{2})}}{4\kappa\sinh^{2}\frac{B}{2}}\\
b=2\mathrm{arcsinh}\sqrt{\frac{\sqrt{64\kappa^{2}+\left(  8\kappa\frac{\mu
}{\left\vert \text{$\Delta$}\right\vert }+1\right)  ^{2}}+8\kappa\frac{\mu
}{\left\vert \text{$\Delta$}\right\vert }+1}{32\kappa^{2}}}%
\end{array}
, \label{EqIng}%
\end{equation}
where $D[J]$ is the domain restricted by the condition Eq.(\ref{EqBorder})
which is explicitly given for the variables $A$ and $B$, as it is shown above.
One\ finds%
\[%
\begin{array}
[c]{c}%
4\int\limits_{-b}^{b}dB\left.  \frac{\arctan\left(  \frac{\left(  2\cosh
\frac{B}{2}-\Omega\right)  \tanh\frac{A}{2}}{\sqrt{4\cosh^{2}\frac{B}%
{2}-\Omega^{2}}}\right)  \cosh^{2}\frac{B}{2}}{\sqrt{4\cosh^{2}\frac{B}%
{2}-\Omega^{2}}}\right\vert _{A=a_{-}}^{A=a_{+}},\\
\int\limits_{-b}^{b}dB\left.  \left(  \frac{\Omega^{2}\arctan\frac
{\Omega+2e^{-A}\cosh\frac{B}{2}}{\sqrt{4\cosh^{2}\frac{B}{2}-\Omega^{2}}}%
}{\sqrt{4\cosh^{2}\left(  \frac{B}{2}\right)  -\Omega^{2}}\cosh^{2}\frac{B}%
{2}}+\frac{e^{-A}}{\cosh\frac{B}{2}}-\Omega\frac{\log\left(  \Omega+2\cosh
A\cosh\frac{B}{2}\right)  -A}{2\cosh^{2}\frac{B}{2}}\right)  \right\vert
_{A=a_{-}}^{A=a_{+}},\\
\int\limits_{-b}^{b}dB\left.  \left(  A+\log\left(  \Omega+2\cosh A\cosh
\frac{B}{2}\right)  -\frac{2\Omega\arctan\frac{\Omega+2e^{A}\cosh\frac{B}{2}%
}{\sqrt{4\cosh^{2}\frac{B}{2}-\Omega^{2}}}}{\sqrt{4\cosh^{2}\frac{B}{2}%
-\Omega^{2}}}\right)  \right\vert _{A=a_{-}}^{A=a_{+}},
\end{array}
\]
for the first, the second and the third integrals Eq.(\ref{EqIng}),
respectively. Integration over $dB$ has to be done numerically. Results of the
numerical calculations are shown in the following figures.%
\begin{center}
\includegraphics[
height=1.9611in,
width=3.1018in
]%
{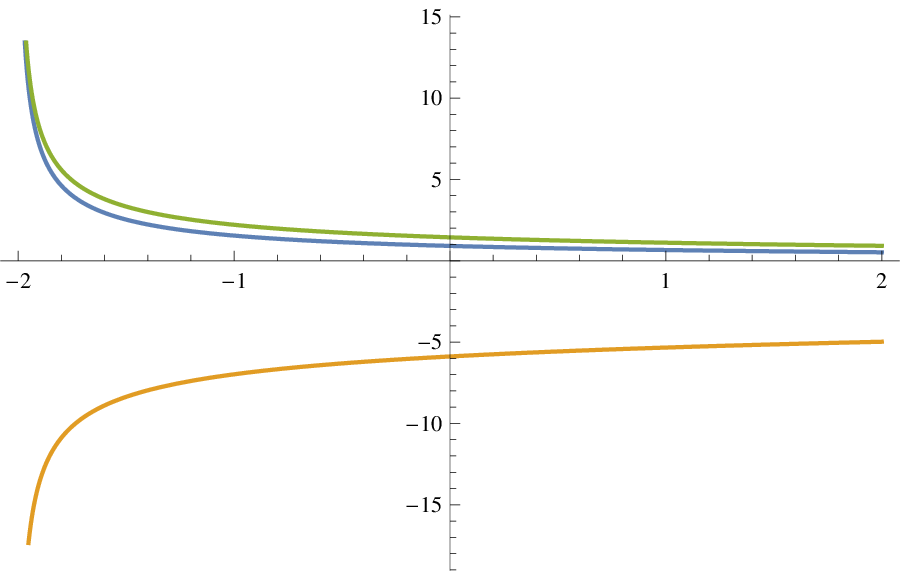}%
\\
Integrals that after symmetrization or antisymmetrization give Eq.(\ref{EqIng}%
) as functions of $\Omega$ in the interval from $\Omega=-2$ till $\Omega=2$.
The parameters are: $\frac{\mu}{\left\vert \text{$\Delta$}\right\vert }%
\simeq580$ and $\kappa\simeq22275$. The integral $\mathcal{I}_{1}\left(
\Omega\right)  $ corresponds to the blue curve, $\mathcal{I}_{2}\left(
\Omega\right)  $ is negative (corresponds to the orange color), and
$\mathcal{I}_{3}\left(  \Omega\right)  $ corresponds to the green curve. The
dependencies on $\frac{\mu}{\left\vert \text{$\Delta$}\right\vert }$ and on
$\kappa$ are very weak, having logarithmic character, shown for $\mathcal{I}%
_{2}\left(  \Omega=0,\frac{\mu}{\left\vert \text{$\Delta$}\right\vert }%
,\kappa\right)  $ in Fig.\ref{FIGI2onPar}.
\label{FIGInt}%
\end{center}
\begin{center}
\includegraphics[
height=2.1128in,
width=3.9409in
]%
{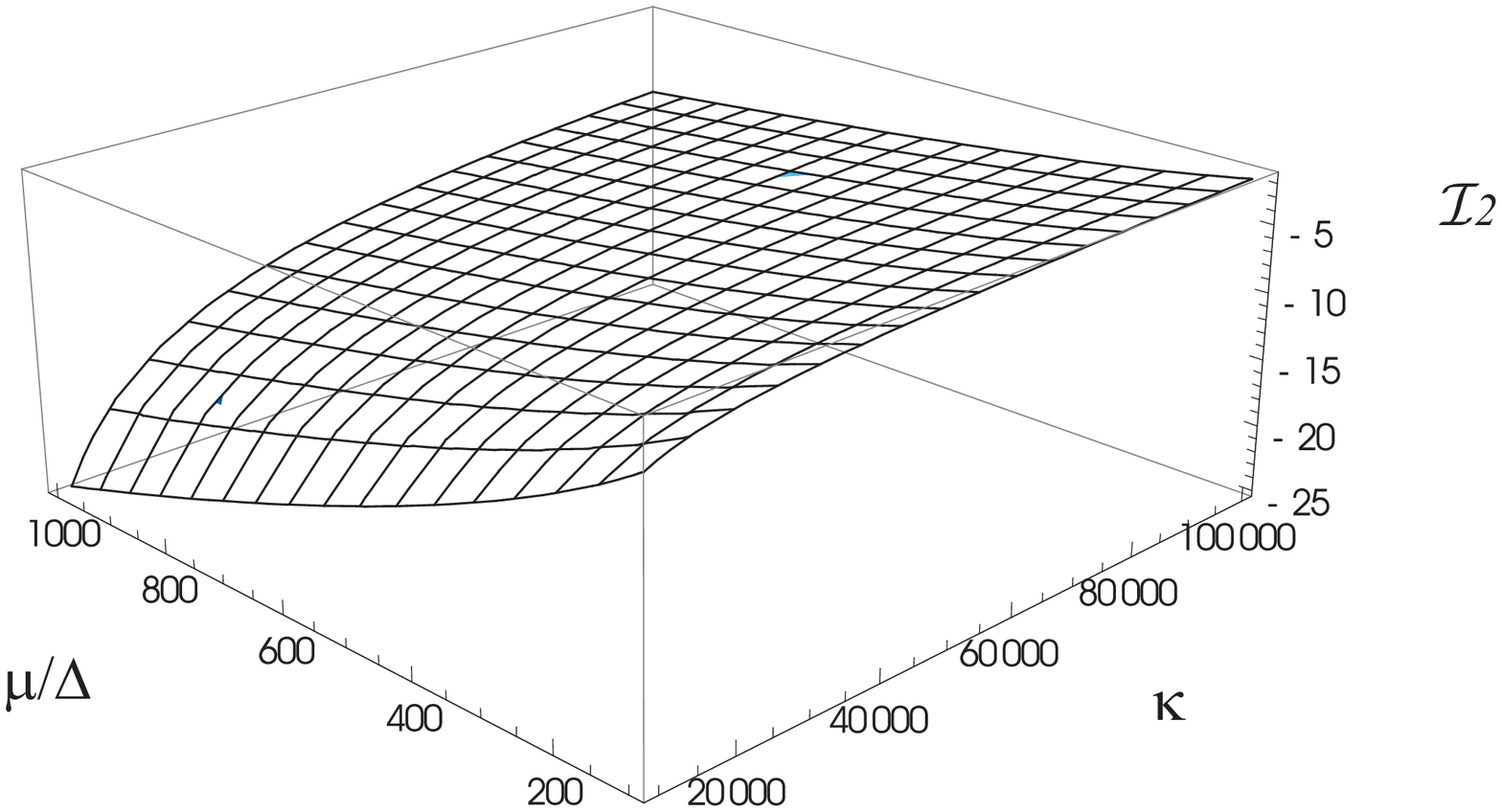}%
\\
Dependence of $\mathcal{I}_{2}\left(  \Omega=0,\frac{\mu}{\left\vert
\text{$\Delta$}\right\vert },\kappa\right)  $\ on the parametrs $\frac{\mu
}{\left\vert \text{$\Delta$}\right\vert }$ and $\kappa$.
\label{FIGI2onPar}%
\end{center}

Combining Eqs.(\ref{EQ35},\ref{EQM},\ref{EQ38},\ref{EQ45},\ref{EqJacFact}%
,\ref{EQTaylor1}) with Eq.(\ref{EQTeylor}), one arrives at%
\begin{align*}
&  \frac{\left\vert \text{$\Delta$}\right\vert i\pi}{8\pi\sqrt{\frac
{\text{$\delta L$}^{2}}{R^{2}}+\text{$\delta k$}^{2}}}\times\\
&  \left(
\begin{array}
[c]{ccc}%
\alpha^{\ast} & \Delta_{1} & \Delta_{2}^{\ast}%
\end{array}
\right)  \left(
\begin{array}
[c]{ccc}%
-2\mathcal{I}_{1}\left(  \Omega\right)  \int\left\vert \overrightarrow{u}%
_{k}^{\ast}\overrightarrow{u}_{k^{\prime}}\right\vert ^{2}n_{e}dV &
\frac{\Delta}{\left\vert \Delta\right\vert }\mathcal{I}_{3}\left(
\Omega\right)  \int\frac{\left(  \overrightarrow{u}_{k}^{\ast}\overrightarrow
{u}_{k^{\prime}}\right)  n_{e}dV}{\sqrt{2\pi dRl}} & -\frac{\Delta^{\ast}%
}{\left\vert \Delta\right\vert }\mathcal{I}_{3}\left(  \Omega\right)
\int\frac{\left(  \overrightarrow{u}_{k}^{\ast}\overrightarrow{u}_{k^{\prime}%
}\right)  n_{e}dV}{\sqrt{2\pi dRl}}\\
\frac{\Delta^{\ast}}{\left\vert \Delta\right\vert }\mathcal{I}_{3}\left(
\Omega\right)  \int\frac{\left(  \overrightarrow{u}_{k^{\prime}}^{\ast
}\overrightarrow{u}_{k}\right)  n_{e}dV}{\sqrt{2\pi dRl}} & \left(
\mathcal{I}_{2}\left(  \Omega\right)  +\mathcal{I}_{4}\left(  \Omega\right)
\right)  \int\frac{n_{e}dV}{2\pi dRl} & \mathcal{I}_{4}\left(  \Omega\right)
\int\frac{n_{e}dV}{2\pi dRl}\frac{\Delta^{\ast2}}{\left\vert \Delta\right\vert
^{2}}\\
-\frac{\Delta}{\left\vert \Delta\right\vert }\mathcal{I}_{3}\left(
\Omega\right)  \int\frac{\left(  \overrightarrow{u}_{k^{\prime}}^{\ast
}\overrightarrow{u}_{k}\right)  n_{e}dV}{\sqrt{2\pi dRl}} & \frac{\Delta^{2}%
}{\left\vert \Delta\right\vert ^{2}}\mathcal{I}_{4}\left(  \Omega\right)
\int\frac{n_{e}dV}{2\pi dRl} & \left(  \mathcal{I}_{2}\left(  \Omega\right)
+\mathcal{I}_{4}\left(  \Omega\right)  \right)  \int\frac{n_{e}dV}{2\pi dRl}%
\end{array}
\right)  \left(
\begin{array}
[c]{c}%
\alpha\\
\Delta_{1}^{\ast}\\
\Delta_{2}%
\end{array}
\right)  .
\end{align*}
After performing the integration over the volume with the allowance for
Eq.(\ref{EQ24}) one finds%
\[
\frac{i\left\vert \text{$\Delta$}\right\vert n_{e}}{8\pi\sqrt{\frac
{\text{$\delta L$}^{2}}{R^{2}}+\text{$\delta k$}^{2}}}\left(
\begin{array}
[c]{ccc}%
\alpha^{\ast} & \Delta_{1} & \Delta_{2}^{\ast}%
\end{array}
\right)  \left(
\begin{array}
[c]{ccc}%
-2\mathcal{I}_{1}\left(  \Omega\right)  O_{p} & \frac{\Delta}{\left\vert
\Delta\right\vert }\mathcal{I}_{3}\left(  \Omega\right)  O_{po} &
-\frac{\Delta^{\ast}}{\left\vert \Delta\right\vert }\mathcal{I}_{3}\left(
\Omega\right)  O_{po}\\
\frac{\Delta^{\ast}}{\left\vert \Delta\right\vert }\mathcal{I}_{3}\left(
\Omega\right)  O_{po}^{\ast} & \mathcal{I}_{2}\left(  \Omega\right)
+\mathcal{I}_{4}\left(  \Omega\right)  & \mathcal{I}_{4}\left(  \Omega\right)
\frac{\Delta^{\ast2}}{\left\vert \Delta\right\vert ^{2}}\\
-\frac{\Delta}{\left\vert \Delta\right\vert }\mathcal{I}_{3}\left(
\Omega\right)  O_{po}^{\ast} & \frac{\Delta^{2}}{\left\vert \Delta\right\vert
^{2}}\mathcal{I}_{4}\left(  \Omega\right)  & \mathcal{I}_{2}\left(
\Omega\right)  +\mathcal{I}_{4}\left(  \Omega\right)
\end{array}
\right)  \left(
\begin{array}
[c]{c}%
\alpha\\
\Delta_{1}^{\ast}\\
\Delta_{2}%
\end{array}
\right)  ,
\]
where%
\begin{align*}
O_{p}  &  =\int\left\vert \overrightarrow{u}_{k}^{\ast}\overrightarrow
{u}_{k^{\prime}}\right\vert ^{2}dV=\frac{\left(  \pi v/c\right)  ^{2}}%
{2\omega_{k^{\prime}}\omega_{k}}\frac{\pi Rd\left(  \overrightarrow{u}%
_{k}^{\ast}\left(  R\right)  \cdot\overrightarrow{u}_{k^{\prime}}\left(
R\right)  \right)  ^{2}}{l}\\
O_{po}  &  =\int\frac{\left(  \overrightarrow{u}_{k}^{\ast}\overrightarrow
{u}_{k^{\prime}}\right)  dV}{\sqrt{2\pi dRl}}=\frac{\pi v/c}{\sqrt{2\omega
_{k}\omega_{k^{\prime}}}}\frac{\sqrt{\pi Rd}\left(  \overrightarrow{u}%
_{k}^{\ast}\left(  R\right)  \cdot\overrightarrow{u}_{k^{\prime}}\left(
R\right)  \right)  }{\sqrt{l}}\\
O_{o}  &  =1
\end{align*}
are the overlap integrals of the mode functions in the domain occupied by the superconductor.

The phases of the unperturbed order parameter and the phase difference of the
field modes can be included to the phases of $\alpha$, $\Delta_{1}$ and
$\Delta_{2}$, which yields%
\[
\frac{i\left\vert \text{$\Delta$}\right\vert n_{e}}{8\pi\sqrt{\frac
{\text{$\delta L$}^{2}}{R^{2}}+\text{$\delta k$}^{2}}}\left(
\begin{array}
[c]{ccc}%
\alpha^{\ast} & \Delta_{1} & \Delta_{2}^{\ast}%
\end{array}
\right)  \left(
\begin{array}
[c]{ccc}%
-2\mathcal{I}_{1}\left(  \Omega\right)  O_{p} & \mathcal{I}_{3}\left(
\Omega\right)  O_{po} & -\mathcal{I}_{3}\left(  \Omega\right)  O_{po}\\
\mathcal{I}_{3}\left(  \Omega\right)  O_{po}^{\ast} & \mathcal{I}_{2}\left(
\Omega\right)  +\mathcal{I}_{4}\left(  \Omega\right)  & \mathcal{I}_{4}\left(
\Omega\right) \\
-\mathcal{I}_{3}\left(  \Omega\right)  O_{po} & \mathcal{I}_{4}\left(
\Omega\right)  & \mathcal{I}_{2}\left(  \Omega\right)  +\mathcal{I}_{4}\left(
\Omega\right)
\end{array}
\right)  \left(
\begin{array}
[c]{c}%
\alpha\\
\Delta_{1}^{\ast}\\
\Delta_{2}%
\end{array}
\right)  .
\]
The action now reads%
\[
i\left(
\begin{array}
[c]{ccc}%
\alpha^{\ast} & \Delta_{1} & \Delta_{2}^{\ast}%
\end{array}
\right)  \left(
\begin{array}
[c]{ccc}%
-2\nu\mathcal{I}_{1}\left(  \Omega\right)  O_{p} & \nu\mathcal{I}_{3}\left(
\Omega\right)  O_{po} & -\nu\mathcal{I}_{3}\left(  \Omega\right)  O_{po}\\
\nu\mathcal{I}_{3}\left(  \Omega\right)  O_{po} & \nu\mathcal{I}_{2}\left(
\Omega\right)  +\nu\mathcal{I}_{4}\left(  \Omega\right)  +\frac{1}{2g} &
\nu\mathcal{I}_{4}\left(  \Omega\right) \\
-\nu\mathcal{I}_{3}\left(  \Omega\right)  O_{po} & \nu\mathcal{I}_{4}\left(
\Omega\right)  & \nu\mathcal{I}_{2}\left(  \Omega\right)  +\nu\mathcal{I}%
_{4}\left(  \Omega\right)  +\frac{1}{2g}%
\end{array}
\right)  \left(
\begin{array}
[c]{c}%
\alpha\\
\Delta_{1}^{\ast}\\
\Delta_{2}%
\end{array}
\right)
\]
where%
\[
\nu=\frac{\left\vert \text{$\Delta$}\right\vert n_{e}}{8\pi\sqrt
{\frac{\text{$\delta L$}^{2}}{R^{2}}+\text{$\delta k$}^{2}}},
\]
and the Gaussian integration over $d\Delta_{1}d\Delta_{2}$ gives%
\[
Z\left(  \alpha^{\ast},\alpha\right)  =\mathrm{const}\exp\left[
\frac{-\left\vert \text{$\Delta$}\right\vert n_{e}\alpha^{\ast}\alpha}%
{4\pi\sqrt{\frac{\text{$\delta L$}^{2}}{R^{2}}+\text{$\delta k$}^{2}}}\left(
O_{p}\mathcal{I}_{1}\left(  \Omega\right)  +\frac{\mathcal{I}_{3}^{2}\left(
\Omega\right)  O_{po}^{2}\left(  1-\delta_{0}^{\delta L}\right)  }{\frac
{4\pi\sqrt{\text{$\delta L$}^{2}/R^{2}+\text{$\delta k$}^{2}}}{g\left\vert
\text{$\Delta$}\right\vert n_{e}}+\mathcal{I}_{2}\left(  \Omega\right)
}\right)  \right]
\]
and after taking the derivative Eq.(\ref{EQ26}) finally yields%
\[
\chi_{k,k^{\prime},\overline{k}^{\prime},\overline{k}}=\frac{-\left\vert
\text{$\Delta$}\right\vert n_{e}}{4\pi\sqrt{\frac{\text{$\delta L$}^{2}}%
{R^{2}}+\text{$\delta k$}^{2}}}\left(  O_{p}\mathcal{I}_{1}\left(
\Omega\right)  +\frac{\mathcal{I}_{3}^{2}\left(  \Omega\right)  O_{po}%
^{2}\left(  1-\delta_{0}^{\delta L}\right)  }{\frac{4\pi\sqrt{\text{$\delta
L$}^{2}/R^{2}+\text{$\delta k$}^{2}}}{g\left\vert \text{$\Delta$}\right\vert
n_{e}}+\mathcal{I}_{2}\left(  \Omega\right)  }\right)  ,
\]
or explicitly%
\begin{equation}
\chi_{k,k^{\prime},\overline{k}^{\prime},\overline{k}}=\frac{-\left\vert
\text{$\Delta$}\right\vert n_{e}\pi^{2}}{8\sqrt{\frac{\text{$\delta L$}^{2}%
}{R^{2}}+\text{$\delta k$}^{2}}}\frac{Rd\left(  \overrightarrow{u}_{k}^{\ast
}\left(  R\right)  \cdot\overrightarrow{u}_{k^{\prime}}\left(  R\right)
\right)  ^{2}}{\left(  c/v\right)  ^{2}\omega_{k^{\prime}}\omega_{k}l}\left(
\mathcal{I}_{1}\left(  \Omega\right)  +\frac{\mathcal{I}_{3}^{2}\left(
\Omega\right)  \left(  1-\delta_{0}^{\delta L}\right)  }{\frac{4\pi
\sqrt{\text{$\delta L$}^{2}/R^{2}+\text{$\delta k$}^{2}}}{g\left\vert
\text{$\Delta$}\right\vert n_{e}}+\mathcal{I}_{2}\left(  \Omega\right)
}\right)  \label{EQFL}%
\end{equation}
Presence of the Kronekker delta $\delta_{0}^{\delta L}$ is due to the fact
that for the case $\delta L=0$, one finds $\Delta_{1}=\Delta_{2}^{\ast}$,
which results in the fact that the collective mode becomes forbidden for the
Raman transition and cannot be excited. Finally, one finds the nonlinear
susceptibility%
\begin{equation}
\chi_{k,k^{\prime},\overline{k}^{\prime},\overline{k}}=\frac{-\left\vert
\text{$\Delta$}\right\vert n_{e}d\left(  \overrightarrow{u}_{k}^{\ast}\left(
R\right)  \cdot\overrightarrow{u}_{k^{\prime}}\left(  R\right)  \right)  ^{2}%
}{32R\sqrt{\frac{\text{$\delta L$}^{2}}{R^{2}}+\text{$\delta k$}^{2}}\left(
c/v\right)  ^{2}\omega_{k^{\prime}}\omega_{k}l}h\left(  \Omega\right)
\label{EQFLL}%
\end{equation}
which couples photons with the wave-vectors $k,k^{\prime},\overline{k}%
^{\prime},\overline{k}$ satisfying the condition $k-k^{\prime}=\overline
{k}-\overline{k}^{\prime}=\delta k$, $L-L^{\prime}=\overline{L}-\overline
{L}^{\prime}=\delta L$. The frequency profile reads%
\begin{equation}
h\left(  \Omega\right)  =\mathcal{I}_{1}\left(  \Omega\right)  +\frac{\left(
1-\delta_{\delta L}^{0}\right)  \mathcal{I}_{3}^{2}\left(  \Omega\right)
}{\frac{4\pi\sqrt{\text{$\delta L$}^{2}/R^{2}+\text{$\delta k$}^{2}}%
}{g\left\vert \text{$\Delta$}\right\vert n_{e}}+\mathcal{I}_{2}\left(
\Omega\right)  }.
\end{equation}
In Eq.(\ref{EQFLL}), in contrast to Eq.(\ref{EQFL}), the factor $\left(
\overrightarrow{u}_{k}^{\ast}\left(  R\right)  \cdot\overrightarrow
{u}_{k^{\prime}}\left(  R\right)  \right)  ^{2}$ has been replaced
by$\frac{\left(  \overrightarrow{u}_{k}^{\ast}\left(  R\right)  \cdot
\overrightarrow{u}_{k^{\prime}}\left(  R\right)  \right)  ^{2}}{\left(  2\pi
R_{inn}\right)  ^{2}}$ to allow for the normalization of the radial mode
functions of the photons.\ With this expression one can substitute the scalar
product $\left(  \overrightarrow{u}_{k}^{\ast}\left(  R\right)  \cdot
\overrightarrow{u}_{k^{\prime}}\left(  R\right)  \right)  =-0.4$ that has been
found earlier for the radius scaled to unity.

The collective mode exists when the equation
\[
\frac{4\pi\sqrt{\text{$\delta L$}^{2}/R^{2}+\text{$\delta k$}^{2}}%
}{g\left\vert \text{$\Delta$}\right\vert n_{e}}+\mathcal{I}_{2}\left(
\Omega\right)  =0
\]
has solutions for $\Omega^{2}$ $<4$ . This happens if $\frac{4\pi
\sqrt{\text{$\delta L$}^{2}/R^{2}+\text{$\delta k$}^{2}}}{g\left\vert
\text{$\Delta$}\right\vert n_{e}}$ $>-\mathcal{I}_{2}\left(  \Omega=0\right)
.\ $However, due to the logarithmic character of the divergency of
$\mathcal{I}_{2}\left(  \Omega\right)  $ at $\Omega^{2}$ $\rightarrow4$, the
collective mode turns out to be exponentially close to the gap boarders, when
the left side of this equality becomes much smaller as compared to the right side.

\section{Appendix \label{Appendix F}}

The collective mode exists when the equation
\[
\frac{4\pi\sqrt{\text{$\delta L$}^{2}/R^{2}+\text{$\delta k$}^{2}}%
}{g\left\vert \text{$\Delta$}\right\vert n_{e}}+\mathcal{I}_{2}\left(
\Omega\right)  =0
\]
has solutions for $\Omega^{2}$ $<4$ . This happens if Eq.(\ref{EQ45bis})
$\frac{4\pi\sqrt{\text{$\delta L$}^{2}/R^{2}+\text{$\delta k$}^{2}}%
}{g\left\vert \text{$\Delta$}\right\vert n_{e}}+\mathcal{I}_{2}\left(
\Omega=0\right)  $ $>0$ holds.$\ $This condition depends on the typical "size
"$1/\sqrt{\text{$\delta L$}^{2}/R^{2}+\text{$\delta k$}^{2}}$ of the waveguide
and on the properties of the superconductors. By employing the relations
\begin{align*}
g\left[  a.u.\right]   &  \simeq\frac{20.045}{n_{e}^{1/3}\left[  a.u.\right]
\log\left(  \frac{861330.n_{e}^{2/3}\left[  a.u.\right]  }{T_{c}\left[  K%
{{}^\circ}%
\right]  }\right)  }\\
\Delta\left[  a.u.\right]   &  \simeq5.555\times10^{-6}T_{c}\left[  K%
{{}^\circ}%
\right] \\
\mu\left[  a.u.\right]   &  \simeq\frac{1}{2}3^{2/3}\pi^{4/3}n_{e}%
^{2/3}\left[  a.u.\right]
\end{align*}
one finds the domains where the collective modes exist and are reasonably far
from the gap borders. This domain depends both on the superconducting material
properties and on its typical size $1/\sqrt{\delta\text{$k^{2}+\delta
L^{2}/R^{2}$}}$ given by the radius $R$ of the tube and the mode wavenumber
difference $\delta k$. In the following figure%
\begin{center}
\includegraphics[
height=2.7771in,
width=3.9834in
]%
{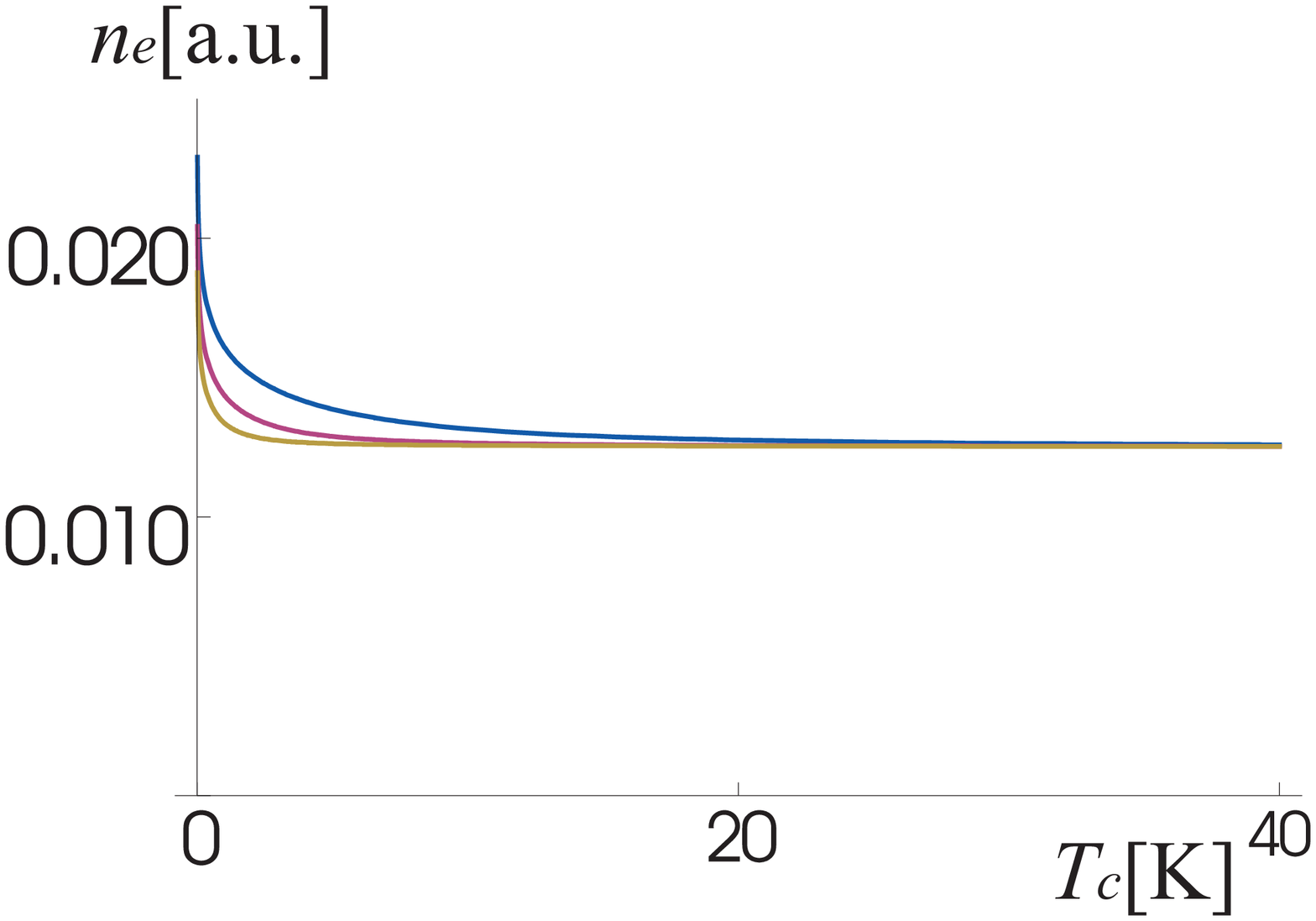}%
\\
Domains of the critical \ temperatures (absis axis) and the electron densities
(ordinate axis, in atomic units $0.01[a.u.]\simeq0.7\times10^{23}cm^{-3}$)
where the collective mode is possible are above the curves. The curves
correspond to typical sizes $1/\sqrt{\delta\text{$k^{2}+\delta L^{2}/R^{2}$}}%
$of the waveguides with $\delta k=0.225$ $\mu ^{-1}$ and $R$ equals to $1$
$\mu$ (top), $3$ $\mu$ (middle), and $10$ $\mu$ (bottom).
\label{FIGBORD}%
\end{center}
one sees this dependence for tube waveguides of radii $1$, $3$, and $10$
microns$\left(
\mu
\right)  $ and for the mode wavenumber difference $\sim0.2%
\mu
^{-1}$.

\section{Appendix \label{Appendix C}}

Since the nonlinear susceptibility for $\delta L=0$ practically does not
depend on the wavevector within the mode, the amplitudes for $m=1,m^{\prime
}=-1$ and for $m=-1,m^{\prime}=1$ in the antisymmetric combination cancel
nonlinearity each of the other in such a way that this combination does not
experience action of the nonlinearity. For the symmetric combination, on the
contrary, the nonlinear coupling given by $\chi_{\delta L=0}l$ acts. The
coupling is local in $z$ and does not depend on the "quantization length" $l$.
By the analogy, the amplitudes for $m=-1,m^{\prime}=-1$ and for $m=1,m^{\prime
}=1$ experience action of the local nonlinear couplings $\chi_{\delta L=-2}l$
and $\chi_{\delta L=2}l$, respectively.

As long as the dependence $k\left(  \omega\right)  $ is restricted to the
linear terms of the Taylor expansions accounting just for the group
velocities, the Schr\"{o}dinger equation for the amplitudes in the coordinate
representation (given by the corresponding Fourier transformation of that in
the momentum representation) belongs to the class of the first order
differential equations
\[
\left(  -i\frac{\partial}{\partial t}-iv\frac{\partial}{\partial z}%
-iv^{\prime}\frac{\partial}{\partial z^{\prime}}+A\left(  z-z^{\prime}\right)
\right)  \Phi(t,z,z^{\prime})=0
\]
in $3$ dimensional space $\left(  t,z,z^{\prime}\right)  $ and therefore can
be solved by the method of characteristics yielding the general solution of
the form
\[
\Phi(t,z,z^{\prime})=\Phi(z-vt,z^{\prime}-v^{\prime}t)e^{-\frac{i}{2}%
\int^{z-z^{\prime}}A\left(  x\right)  dx},
\]
where $\overline{\Phi}(x,y)$ is an arbitrary function of two variables has to
be found from the initial conditions. For two independent bell-shaped accident
wave packets $\overline{\Phi}(x,y)=\phi\left(  x\right)  \overline{\phi
}\left(  y\right)  $ with no initial overlap, that is $\int\phi\left(
z-vt\right)  \overline{\phi}\left(  z-v^{\prime}t\right)  dz=0$ for $t<t_{in}%
$, the asymptotic solution for $t\rightarrow\infty$ reads
\[
\Phi(t,z,z^{\prime})=\phi\left(  z-vt\right)  \overline{\phi}\left(
z^{\prime}-v^{\prime}t\right)  e^{\frac{-i}{2}\int_{-\infty}^{\infty}A\left(
t\delta v\right)  dt}.
\]
For the interaction independent on the wavevector, the function $A\left(
z-z^{\prime}\right)  $ is local, that is proportional to the Dirac delta
function $A\delta_{z-z^{\prime}}$, and the asymptotic form attains after a
finite interval of time $t>t_{fin}$ when the faster wavepacket completely
overtakes the slower one, such that%
\[
\int\phi\left(  z-vt_{fin}\right)  \overline{\phi}\left(  z-v^{\prime}%
t_{fin}\right)  dz=0.
\]
%

\end{widetext}%


\begin{thebibliography}{99}                                                                                               %


\bibitem {Nakanishi}A. Nakanishi H. Katayama-Yoshida, Solid State
Communications, \textbf{152}, 24-27 (2012)

\bibitem {Exp}A.Kumatani,T.Ohsawa, R.\ Shimizu, Y.\ Takagi,S.\ Shiaki,
T.\ Hitosugi, Appl.\ Phys.Lett,\textbf{101,123103 (2012).}

\bibitem {Dicke}R.H.\ Dicke, Phys.\ Rev. \textbf{93}, 99-110 (1954)

\bibitem {Abrikosov}A.A. Abrikosov, L.P. Gorkov, I.E. Dzyaloshinski,
\textit{Methods of Quantum Field Theory in Statistical Physics}, Richard
A.\ Silverman, ISBN-10:0-486-63228-8, pp. 315-320, (1963)

\bibitem {Kulik}I. O. Kulik O. Entin-Wohlman, R. Orbach, Journal of Low
Temperature Physics \textbf{43}, Issue 5--6, pp 591--620, (1981)

\bibitem {Klein}M.V.\ Klein, S.B. Dierker, Phys.Rev.\textbf{B} 29, 4976-4991 (1984)

\bibitem {UFN}For details one can consult, for instance, the review by I. N.
Toptygin, \ Physics--Uspekhi, \textbf{60}, 935--947 (2017)

\bibitem {MichaFedorov}J.H.\ Eberly, J.\ Javanainen, K.\ Rzazevski, Physics
Reports, \textbf{204},331-383 (1991)

\bibitem {Kleinert}For the details see the book by H.Kleinert,
\textit{Collective Classical and Quantum Fields} \ World Scientific 2018,
ISBN: 978-981-3223-93-6, Sec.\ 3.1

\bibitem {Expr}Eq.(\ref{EQ28}) is the Fourier representation of Eqs.(3.45-47)
of \cite{Kleinert} obtrained from the action$\int dt\mathrm{Tr} $
$\overrightarrow{\psi}^{+}\left(  i\partial_{t}-\widehat{H}\right)
\overrightarrow{\psi}$ with the help of anticommutation and integration in parts.

\bibitem {Leggett}A.J.\ Leggett, Progress of Theoretical Physics, \textbf{36},
901-930 (1966)
\end{thebibliography}
\end{document}